\documentclass[twocolumn]{revtex4-2}
\usepackage{graphicx} 
\usepackage{natbib}
\usepackage{amsmath}
\usepackage{xcolor}
\usepackage{newtxtext}
\usepackage{newtxmath}
\usepackage{orcidlink}
\usepackage{ulem}
\makeatother

\begin{document}
\title{Numerical simulations of waves and turbulence in coronal loops: observables and spectra}
\date{}

\author{F. Feraco$^{1}$\orcidlink{0000-0002-5550-2070}
  ,
    F. Pucci$^{2}$\orcidlink{0000-0002-5272-5404} 
    ,
    C. Meringolo$^{3}$\orcidlink{0000-0001-8694-3058}
    ,
        G. Nistic\`o$^{1,4}$\orcidlink{0000-0003-2566-2820} 
   ,
    F. Reale$^{5,6}$\orcidlink{0000-0002-1820-4824} 
    ,
    P. Pagano$^{5,6}$\orcidlink{0000-0001-5274-515X} 
    ,
    G. Cozzo$^{7}$ \orcidlink{0009-0005-4545-1010} 
    ,
    T. Van~Doorsselaere $^{8}$\orcidlink{0000-0001-9628-4113} 
    , 
    B. De Pontieu$^{9,10,11}$\orcidlink{0000-0002-8370-952X} 
    ,
    P. Testa $^{7}$\orcidlink{0000-0002-0405-0668} 
    ,
    S. Servidio$^{1,2,4}$ \orcidlink{0000-0001-8184-2151} 
    ,
    O. Pezzi$^{2}$ \orcidlink{0000-0002-7638-1706} 
    ,
    F. Valentini$^{1,4}$\orcidlink{0000-0002-1296-1971}
    ,
    F. Malara$^{1,2,4}$\orcidlink{0000-0002-5554-8765} 
    }

   \affiliation{\textsuperscript{1} Dipartimento di Fisica, Università della Calabria, via P. Bucci, 87036 Rende (CS), Italy, 
   \\
        \textsuperscript{2} Istituto per la Scienza e Tecnologia dei Plasmi, Consiglio Nazionale delle Ricerche (CNR-ISTP), 70126 Bari, Italy,
    \\
        \textsuperscript{3}Institut für Theoretische Physik, Goethe Universität, Max-von-Laue-Str. 1, D-60438 Frankfurt am Main, Germany,
    \\
         \textsuperscript{4}Istituto Nazionale di Astrofisica - INAF, Direzione Scientifica, Viale del Parco Mellini 84, 00136 Roma, Italy,
    \\
       \textsuperscript{5}Dipartimento di Fisica \& Chimica, Università di Palermo, via Archirafi 36, 90134 Palermo, Italy,
    \\
       \textsuperscript{6}National Institute for Astrophysics, Astronomical Observatory of Palermo, Piazza del Parlamento 1, I-90134 Palermo, Italy,
    \\
    \textsuperscript{7}Harvard-Smithsonian Center for Astrophysics, 60 Garden St, Cambridge, MA 02193, USA
    \\
    \textsuperscript{8}Centre for mathematical Plasma Astrophysics, Mathematics Department, KU Leuven, Celestijnenlaan 200B bus 2400, B-3001 Leuven, Belgium
    \\
    \textsuperscript{9}Lockheed Martin Solar \& Astrophysics Laboratory, 3251 Hanover Street, B203, Palo Alto, CA 94304
    \\
    \textsuperscript{10}Rosseland Centre for Solar Physics, University of Oslo, P.O. Box 1029 Blindern, NO-0315, Oslo, Norway
    \\
    \textsuperscript{11}Institute of Theoretical Astrophysics, University of Oslo, P.O. Box 1029 Blindern, NO-0315, Oslo, Norway
 }

\begin{abstract}
Coronal loops are often regarded as the building blocks of the solar corona, however the heating mechanism and complex dynamics resulting from the turbulent nature of the plasma and the presence of waves is still unclear.
We  investigate numerically the time evolution of velocity and magnetic field fluctuations in a coronal loop, focusing on the dynamics due to both phase mixing and turbulent cascade. The intensity, doppler velocity and non-thermal broadening are synthesized from numerical results in order to establish if the upcoming Multi-slit Solar Explorer (MUSE) mission could reveal the presence of those phenomena in the solar corona through its unprecedented high-resolution spectroscopic observations.
The time evolution of velocity and magnetic field fluctuations in coronal loops is studied by means of numerical three-dimensional (3D) magnetohydrodynamics (MHD) simulations. The loop is represented by a cylindrical pressure-balanced magnetic structure with a transverse density and magnetic field inhomogeneity. The initial perturbation is a superposition of a torsional Alfvén wave and a transverse turbulent component with different tunable weights. Heating effects due to viscous and resistive dissipation are taken into account. In order to reconstruct plasma emission features we calculate moments of the Fe IX 171 \AA\ spectral line, synthesized from the simulations. 2D maps obtained by integrating the emission along the assumed line of sight are calculated for the emission intensity $I_0$, the Doppler shift $I_1$ and the non-thermal broadening $I_2$ as functions of time, for several values of the model parameters. Finally, we simulate MUSE spectrograph by considering a resolution of $312$ km $\times$ $312$ km.
We observe how intensity maps show the formation of longitudinal threads during time evolution. The generation of small-scale fluctuations mainly takes place in the inhomogeneity region at the loop boundary, where the effects of phase mixing and non-thermal broadening are stronger.
In these regions the presence of transverse oscillations is clearly visible in $I_1$ and $I_2$ maps.
1D power spectra of intensity and Doppler shift maps are calculated and compared with the corresponding spectra of density and line-of-sight velocity component. 
At large spatial scales (i.e. between $300$ km and $3000$ km) the spectral indexes of density and intensity spectra are very similar, while the spectrum of $I_1$ is steeper.
These results are discussed in the perspective of a forward modeling of observations of fluctuation dynamics in coronal loops, in particular by MUSE spectrograph.
The resolution attained by MUSE will allow to obtain power spectra of coronal loops with a resolution never achieved before.
The agreement observed between the spectral indexes of the intensity power spectra at MUSE resolution and the one computed from the full 3D density field indicates that spectra of $I_0$ can be used to infer information on the spectrum of density inside a loop.
\end{abstract}

\maketitle

\section{Introduction}
Waves and oscillations in the magnetic structures of the solar corona have received a remarkable attention in the last decades, both from theoretical and observational point of view. Dissipation of the energy associated with waves is considered a possible mechanism to explain the heating of the coronal plasma \citep{nakariakov2020magnetohydrodynamic}. Moreover, plasma and magnetic field fluctuations, which are detected in-situ in the solar wind (e.g. \cite{Bruno13}) and are partially responsible of wind heating and acceleration \citep{marinosorriso_23, Sorriso-Valvo24}, probably have a coronal origin \citep{rivera_24}.\\
\noindent One of the possible sources of non-thermal broadening are large-scale oscillations in the coronal plasma \citep{Feldman88,Dere93}.
Both slow magnetohydrodynamic (MHD) waves \citep{Chae98,Ofman99} and transverse oscillations in post-flare loops \citep{Aschwanden99,Nakariakov99,Schrijver99} have been detected in coronal magnetic structures. Standing kink oscillations in coronal loops have been observed \cite[e.g.,][]{Zimovets15,Goddard16}, which are subject to damping within few wave periods and could be excited by coronal impulsive events (see \citet{Nakariakov21} for a review). Similarly, decayless oscillations have been detected mainly in quiet loops with no apparent damping during several periods \citep{Tian12,Wang12,Nistico13} and interpreted as standing waves \citep{Anfinogentov15}. Both the interaction between loops and quasi-steady flows \citep{Nakariakov16}, and continuous footpoint driving \citep{Karampelas17,Karampelas19,Afanasyev20} have been proposed as mechanisms responsible for such decayless oscillations.
In addition, propagating, mainly transverse waves have been revealed in the corona both by ground-based (CoMP, \citep{Tomczyk07,Tomczyk09}) and by space-based (AIA, on board SDO \citep{depontieu_07,McIntosh11}) observations. These oscillations have been interpreted either as (mainly) outward-propagating Alfvén waves, or as fast-mode kink waves in cylindrical structures \citep{VanDoorsselaere08}. Their ubiquity seems to indicate a possible role of such waves in coronal heating.\\
The presence of space inhomogeneities in coronal structures strongly influences the dynamics of waves \citep[e.g.,][]{pucci_16}. For instance, variations of the background density in the direction transverse to the magnetic field generate small scales in perturbations through a variety of mechanisms. Such inhomogeneities represent a crucial element in many wave-based theories of coronal heating. In fact, in the high-Reynolds/Lundquist-number coronal plasma an effective conversion of the wave energy into heat requires a fast generation of small scales \citep{terradas2018}. 
In this framework, phase mixing of Alfvén waves has been considered, which is due to transverse gradients of the Alfvén velocity \citep{Heyvaerts83,McLaughlin2011}. Differences in the propagation velocity in nearby magnetic lines progressively bend wave fronts, thus generating smaller and smaller scales in the perturbation in the direction transverse to the magnetic field. Phase mixing has been studied using different techniques, normal-mode \citep{Steinolfson85,Califano90,Califano92} or initial-value \citep{Lee86,Malara92,Malara96} approaches and in a variety of situations, such as longitudinal variations of density or magnetic field \citep{Ruderman98}, 3D magnetic structures in the small wavelength limit \citep{Petkaki98,Malara00}, models of quiet-Sun \citep{Malara03,Malara05,Malara07} or open field line coronal regions \citep{Malara13,Pucci14}. Moreover, kink modes initially present in inhomogeneous loops can resonantly transfer their energy to Alfvén modes through a process known as resonant absorption \citep{Terradas08} which are then subject to phase mixing \citep{goossens_92,pagano_18,pagano_19,pagano_20}. Phase mixing in complex coronal plasma has been also studied \citep{howson2020phase} showing how an enhanced magnetic complexity leads to a faster wave dissipation.\\
Torsional Alfvén waves are non-compressive, azimuthally polarized oscillations propagating in cylindrically-symmetric structure. Such kind of oscillations could be generated by torsional motions at the loop bases. Rotation motions have been detected both in the photosphere \citep{Brant88,Bonet08,Bonet10} and in the chromosphere \citep{Wedemeyer09,DePontieu_12,DePontieu_14,Tziotziou18,Tziotziou20,Dakanalis22}, as well as in the transition region \citep{Wedemeyer12,DePontieu_14} or in magnetic funnels \citep{Jess09}. The time evolution of torsional Alfvén waves is mainly determined by phase mixing, making those oscillations the simplest case for investigating phase mixing in cylindrically symmetric structures. However, for large enough wave amplitudes, strong velocity shears induced by phase mixing lead to a Kelvin-Helmholtz instability which, in turn, generates a turbulent state in the inhomogeneity region \citep{diaz2021transition,diaz2022transition}.\\
In a turbulent fluid small scales are generated by nonlinear couplings among perturbations at different wavelengths. This generates an energy cascade toward small scales, which leads to the formation of a power-law spectrum.
While in homogeneous isotropic turbulence the energy is transferred from large to small scales, in the presence of waves and/or strong anisotropy, inverse or bi-directional cascades can also develop \citep{pouquet_13,marino_15}, as observed in geophysical flows \citep{balwada_22,alexakis_24} and in the solar wind plasma \citep{marino_25}. 
In the case of alfvénic MHD, nonlinear couplings take place between counter-propagating fluctuations and small scales in the directions perpendicular to the background magnetic field are preferentially formed \citep{shebalin83,Carbone90,Oughton94}. This scenario can be envisioned for closed magnetic structures, as loops. However, even in open-field regions, where upward-propagating fluctuations dominate, partial reflection due to vertical inhomogeneities can generate downward-propagating fluctuations with a consequent formation of a turbulent cascade \citep{pascoe_22,kumar_25}. Non-thermal broadening observed in the corona can be interpreted as due to turbulence \citep{Banerjee98,Singh06,Hahn13,Hahn14}. Moreover, $f^{-1}$ frequency spectra observations \citep{Tomczyk07,Morton16,Morton19} in loops are indicative of turbulent fluctuations. A more direct evidence of a turbulent corona comes from recent observations performed by the Parker Solar Probe spacecraft in the most external coronal layers \citep{Kasper21,Bandyopadhyay22,Zhao22}.\\
Turbulence-based models of coronal heating have been presented, both for coronal loops \citep{vanBallegooijen11,Downs16,rappazzo2017,vanBallegooijen17,vanBallegooijen18} and for open-field regions \citep{Verdini09,Perez13,Woolsey15,Chandran19}. In addition, simplified turbulence model based on reduced MHD have been applied for the heating of coronal loops \citep{Nigro04,Nigro05,Reale05,Buchlin07} and to characterize dissipation \citep{Malara10}. Such models reproduce frequency spectra \citep{Nigro20} with properties similar to those observed in the corona \citep{Tomczyk09,McIntosh11,Morton16}.
Moreover, the effects of wave heating in coronal loops have been studied using MHD numerical simulations with \citep{karampelas_19} and without gravity \citep{shi_21}.\\
Both phase mixing and turbulence contribute to the generation of small scales and promote dissipation. In some situations these two effects can coexist. For instance, Alfvén waves localized at the interface between loop interior and exterior and excited by a kink oscillation, initially undergo phase mixing. Subsequently, a Kelvin-Helmholtz instability comes into play, which drives localized turbulence. This eventually leads to energy dissipation. This mechanism has been investigated in depth \citep{Terradas08,Antolin14,Magyar15,Antolin19} also in comparison with observations \citep{Antolin17, okamoto_15}. Moreover, when phase mixing takes place in a plasma with complex spatial inhomogeneities, a phenomenology similar to that of turbulence can be found \citep{magyar2017generalized}. In a recent paper \citep[][hereafter Paper 1]{meringolo_24} the interplay between phase mixing and turbulence has been studied in inhomogeneous flux tubes modeling a coronal loop. It has been found that, similar as for torsional waves, a form of phase mixing also acts on a turbulent transverse perturbation in the inhomogeneity region. For perturbation amplitudes sufficiently low, phase mixing and turbulence work synergically, speeding up the energy dissipation. In particular, the dissipative time is shorter than those of phase mixing and of the nonlinear cascade, when individually considered.\\
In the present paper we will consider the same configuration as in Paper 1, namely, a superposition of a torsional Alfvén wave and a turbulent transverse perturbation with tunable amplitudes, both propagating in a magnetic flux tube with a transverse inhomogeneity. Our main purpose is calculating maps of observable quantities, namely emission intensity, Doppler shift and non-thermal broadening, relative to the emission from a given spectral line. These quantities will be synthesized from the results of numerical simulations during the time evolution of the perturbation and for different values of the parameters of the model using two different resolutions. The aim is to investigate the actual observability of the considered wave and turbulence phenomena by spectrometers such as those that will be employed for the forthcoming Multi-slit Solar Explorer (MUSE) mission.\\
\noindent With the aim of better understanding the solar corona and the mechanism behind coronal heating, MUSE mission is due to begin no earlier than in 2027 \citep{depontieu_20,depontieu_22}.
Because of its high-cadence spectral raster scans over a large field of view, MUSE will provide constraints for models in 4 spectral lines formed at coronal temperatures in the passbands centered Fe XIX 108 \AA, Fe XXI 108 \AA, Fe IX 171 \AA, and Fe XV 284 \AA\ spectral lines, which are formed at log T of $7.0$, $7.1$, $5.9$, and $6.4$.
The nominal spatial resolution is $0.4$ arcsec, temporal cadence $1-4$ s, spectral resolution $\sim 5$ km/s \citet{depontieu_20}.
In addition, power spectra of intensity and Doppler shift maps will be calculated and compared with density and velocity field power spectra respectively, in order to check whether reliable information about density or the distribution of kinetic energy among spatial scales within the corona can be inferred from an analysis of intensity and Doppler shift maps.

\section{Numerical Framework}

\subsection{MHD equations and numerical method}\label{Subsec.equations}
To describe the nonlinear dynamics of coronal loops, we numerically solve the 3D compressible viscous-resistive MHD equations, written in the following dimensionless form:
\begin{equation}\label{MHD1}
\frac{\partial \rho}{\partial t} + \nabla \cdot \left( \rho \mathbf{v} \right)=0,
\end{equation}
\begin{equation}\label{MHD2}
\frac{\partial \mathbf{v}}{\partial t} + \left( \mathbf{v} \cdot \nabla \right) \mathbf{v} = -\frac{\beta}{2\rho} \nabla \left(\rho T\right) + 
\frac{1}{\rho} \left[ \left( \nabla \times \mathbf{B} \right) \times \mathbf{B} \right] + \frac{1}{\rho} \nabla \cdot \sigma,
\end{equation}
\begin{equation}\label{MHD3}
\frac{\partial \mathbf{A}}{\partial t} = \mathbf{v} \times \mathbf{B} + \eta \nabla^2 \mathbf{A},
\end{equation}
\begin{equation}\label{MHD4}
\begin{aligned}
\frac{\partial T}{\partial t} + \left( \mathbf{v}\cdot \nabla \right) T + \left( \gamma -1\right) T \left( \nabla \cdot \mathbf{v}\right) = \\ \frac{\kappa}{\rho} \nabla^2 T +
\frac{2(\gamma -1)}{\beta \rho} \left[ \eta \left( \nabla \times \mathbf{B}\right)^2 + \frac{1}{2} \sigma : \sigma \right],
\end{aligned}
\end{equation}
where $\rho$ is the mass density; $\mathbf{v}$ is the velocity; $T$ is the temperature; $\mathbf{A}$ is the vector potential; and $\mathbf{B}=\nabla \times \mathbf{A} + \langle \mathbf{B} \rangle$ is the magnetic field, where angular brackets indicate a spatial average over the considered spatial domain $D$. Due to periodic boundary conditions (see below), it is $\langle \nabla \times \mathbf{A} \rangle =0$. Moreover, the term $\langle \mathbf{B} \rangle$, which is spatially uniform by definition, remains constant in time.
In the present simplified periodic model gravity is not included. We plan to add both gravity and radiative energy losses in a future, more realistic model. 
The above quantities are dimensionless: $\mathbf{B}$ is normalized to a typical magnetic field $\tilde{B}$; $\rho$ is normalized to a typical mass density $\tilde{\rho}$; $\mathbf{v}$ is normalized to a typical Alfvén velocity $\tilde{c}_A=\tilde{B}/\sqrt{4\pi \tilde{\rho}}$; and $T$ is normalized to a typical temperature $\tilde{T}$. Spatial coordinates are normalized to a typical length $\tilde{L}$; time is normalized to the Alfvén time $\tilde{t}_A=\tilde{L}/\tilde{c}_A$; $\mathbf{A}$ is normalized to $\tilde{B}\tilde{L}$; and $\gamma$ is the adiabatic index, assumed to be $\gamma=5/3$. 
The values of quantities $\tilde{B}$, $\tilde{\rho}$, $\tilde{T}$ and $\tilde{L}$ will be given in Sect. \ref{Sec:simparam}.
The constant coefficient $\beta$ gives a typical value for the plasma beta. It is defined as the ratio between the typical plasma and magnetic pressures: $\beta=8\pi \kappa_{\rm B} \tilde{\rho}\tilde{T}/(\mu m_{\rm p}\tilde{B}^2)$, with $\kappa_{\rm B}$ the Boltzmann constant, $\mu$ the mean atomic weight (for hydrogen plasma $\mu=1$) and $m_{p}$ the proton mass. In the coronal plasma, it is typically $\beta \ll 1$; we used the value $\beta = 0.01$. In code units, the plasma pressure is given by $P = \beta \rho T/2$. 
The viscous stress tensor $\sigma$ appearing 
in Eq.s (\ref{MHD2}) and (\ref{MHD4}) has components: 
\begin{equation}\label{sigma}
\sigma_{ij} = \nu \left( \frac{\partial v_i}{\partial x_j} + \frac{\partial v_j}{\partial x_i} \right) + \left( \zeta - \frac{2}{3} \nu\right) \left( \sum_{k=x,y,z} \frac{\partial v_k}{\partial x_k} \right) \delta_{ij} 
\;\; , \;\; i,j = x,y,z
\end{equation}
where $\delta_{ij}$ is the Kroneker's delta.
The quantities $\nu=\hat{\nu}/(\tilde{L}\tilde{\rho}\tilde{c}_A)$ and $\zeta=\hat{\zeta}/(\tilde{L}\tilde{\rho}\tilde{c}_A)$ are the normalized viscosity coefficients, $\eta=c^2\hat{\eta}/(4\pi\tilde{L}\tilde{c}_A)$ is the normalized resistivity and $\kappa=2\hat{\kappa}\tilde{T}/(\beta \tilde{L}\tilde{\rho}\tilde{c}_A^3)$ is the normalized thermal diffusivity coefficient, where $\hat{\nu}$ and $\hat{\zeta}$ are the two dynamic viscosity coefficients, $\hat{\eta}$ is the resistivity, $\hat{\kappa}$ is the thermal diffusivity coefficient, and $c$ is the speed of light. 
We set $\nu=\eta=\kappa=\zeta=10^{-5}$ for all the runs but run $B2$, for which $\nu=\zeta=2\cdot10^{-5}$ in order to avoid numerical instabilities due to the larger amplitude. This choice depends on the numerical spatial resolution and in our case allows us to prevent numerical instabilities.
Compressibility is included in the model, in order to describe possible couplings among incompressible and compressible fluctuations, which can enhance the production of small scales \citep[e.g.,][]{Malara96}. In Paper 1 an adiabatic energy equation had been used, and hyper-dissipative terms had been introduced to obtain a wider inertial range in the fluctuation spectra. In contrast, here we are interested in describing plasma heating due to dissipation of fluctuations.
Therefore, we included heating terms in the energy equation (\ref{MHD4}) and adopted standard viscosity, resistivity, and thermal diffusivity terms in Eq.s (\ref{MHD2})-(\ref{MHD4}). 
Dissipation in the coronal plasma is probably due to kinetic mechanisms and takes place at scales much smaller than those considered in the present numerical model. Therefore the values we used for normalized dissipative coefficients are as small as possible, compatibly with the finite spatial resolution of simulations.
Given the small perturbation amplitudes we used for our runs we do not include radiative losses, since their contribution to energy budget would have been negligible.
Finally, to obtain a more accurate description of possible shocks and discontinuities, a logarithmic regularization to the density is used, setting $\rho=e^g$ and solving the equivalent equation for the quantity $g$.\\ 
Equations (\ref{MHD1})–(\ref{MHD4}) are numerically solved in a 3D Cartesian domain with periodicity boundary conditions along the three spatial directions $x$, $y$, and $z$. In particular, $z$ represents the direction of the background magnetic field $\mathbf{B}_0$. The domain $D$ in normalized units is defined as $D=\left\{ (x, y, z)\right\} = \left[ 0, L_\perp \right] \times \left[ 0, L_\perp \right] \times \left[ 0, L_{||}\right]$, with $L_\perp = 2\pi$ and $L_{||}=\Lambda L_\perp$, where $\Lambda$ is a free parameter that represents the aspect ratio of the domain. All quantities are evaluated on a spatial grid formed by $N_\perp \times N_\perp \times N_{||} = 512 \times 512 \times 128$ points along the respective directions $x$, $y$ and $z$.
To calculate the solution we used the COmpressible Hall Magnetohydrodynamics simulator for Plasma Astrophysics (COHMPA) algorithm \citep{Pezzi2024} with the Hall term switched off. COHMPA adopts a pseudo-spectral method to treat spatial dependence and a second-order Runge-Kutta algorithm to advance variables in time. A spectral filter and a $2/3$ rule are employed in the spectral space to suppress numerical artifacts due to aliasing \citep{Orszag1971,Meringolo2021}.  COHMPA has already been used both in 2D \citep{vasconez2015,perri2017,pezzi2017revisiting} and in 3D \citep{Pezzi2024} configurations.

\subsection{Initial condition}
The initial condition is a superposition of an ideal MHD equilibrium and a perturbation. These quantities have the same form as in Paper 1.
Coronal loops are arch-like structures of confined plasma characterized by a higher density compared to the surrounding plasma.
It is observed that the physical characteristics of coronal loops (i.e. length, temperature, density) depend mainly on which region of the Sun they come from \citep{verwichte_13,reale_14,li_20}, and even from the same region the observed values can vary by a factor of $2-3$ \citep{maccormack_19}.
For instance, typical values of the length for coronal loops developing in an active region range from $10$ to $100$ Mm \citep{reale_14}.
The equilibrium is a simplified model of a coronal loop in the form of a cylindrically-symmetric straight magnetic flux tube. The equilibrium magnetic field $\mathbf{B}_0=B_0(r) \hat{\mathbf{z}}$ is directed along $z$, where $\hat{\mathbf{z}}$ is the unit vector in the $z$ direction, $r=\left[ (x-x_0)^2 + (y-y_0)^2 \right]^{1/2}$ is the radial distance from the symmetry axis, with $(x_0,y_0)=(\pi,\pi)$ the coordinates of the symmetry axis in the $xy$ plane. The equilibrium pressure $P_0(r)$ increases toward the loop axis, with a sharp but continuous transition between the loop exterior and interior, of the form:
\begin{equation}\label{P0}
P_0(r)= \frac{(P_{\rm int}-P_{\rm ext})}{2} \left[ 1 - \tanh \left( \frac{r-r_0}{\Delta r}\right)\right] + P_{\rm ext}
\end{equation}
where $P_{\rm ext}$ and $P_{\rm int}$ are the pressure values in the exterior and interior of the loop, respectively, $r_0=\pi/2=L_\perp/4$ is the radius of the cylindrical surface where $P_0(r_0)=(P_{\rm int}+P_{\rm ext})/2$, and $\Delta r=\pi/8=L_\perp/16$ is the thickness of the loop boundary. The equilibrium temperature is assumed to be uniform: $T=T_0=1$. The equilibrium density $\rho_0(r)=2P_0(r)/(\beta T_0)$ varies between $\rho_{\rm ext}=1$ and $\rho_{\rm int}=2$ in the exterior and interior of the loop, respectively. Correspondingly, we have $P_{\rm ext}=\beta \rho_{\rm ext} T_0/2=5\cdot 10^{-3}$ and $P_{\rm int}=\beta \rho_{\rm int} T_0/2= 10^{-2}$.\\
The equilibrium magnetic field profile $B_0(r)$ is calculated by imposing total (plasma + magnetic) pressure equilibrium and $B_0=1$ in the loop exterior. Due to the small value of $\beta$, the pressure equilibrium is achieved with an almost uniform $B_0(r)$ throughout the spatial domain. Finally, the equilibrium velocity is $\mathbf{v}_0=0$. 

\begin{figure}[h]
\centering
\includegraphics[width=0.45\textwidth]{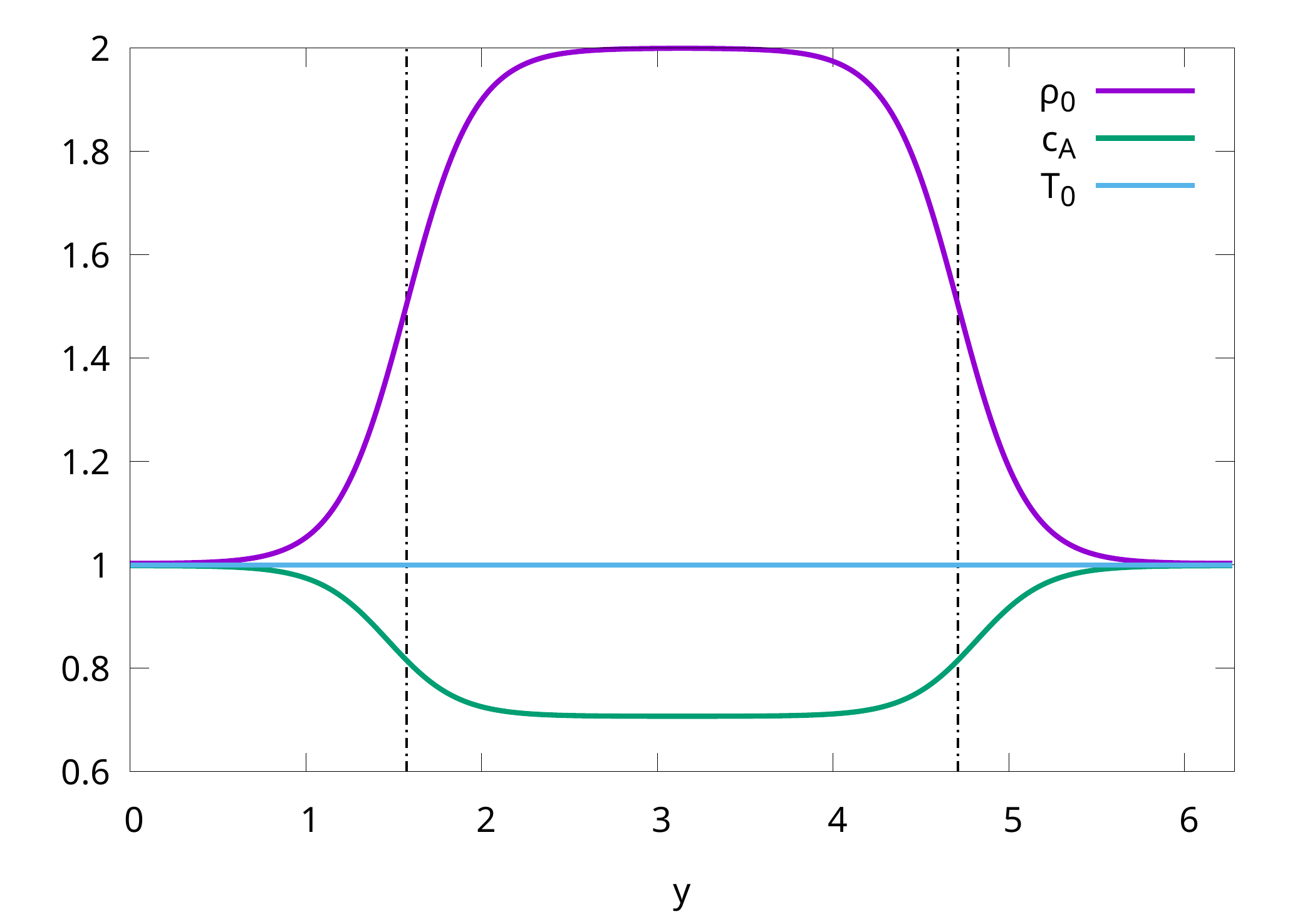}  
\caption{1D profile of density $\rho$, temperature $T$ and Alfvén speed $c_A$ taken at the initial condition across the axis of the loop. The vertical dashed line indicate the loop boundaries, which we conventionally fix at locations where $\rho=1.5$.}
\label{f:profile0.5}
\end{figure} 

\begin{figure}[h]
\centering
\includegraphics[width=0.45\textwidth]{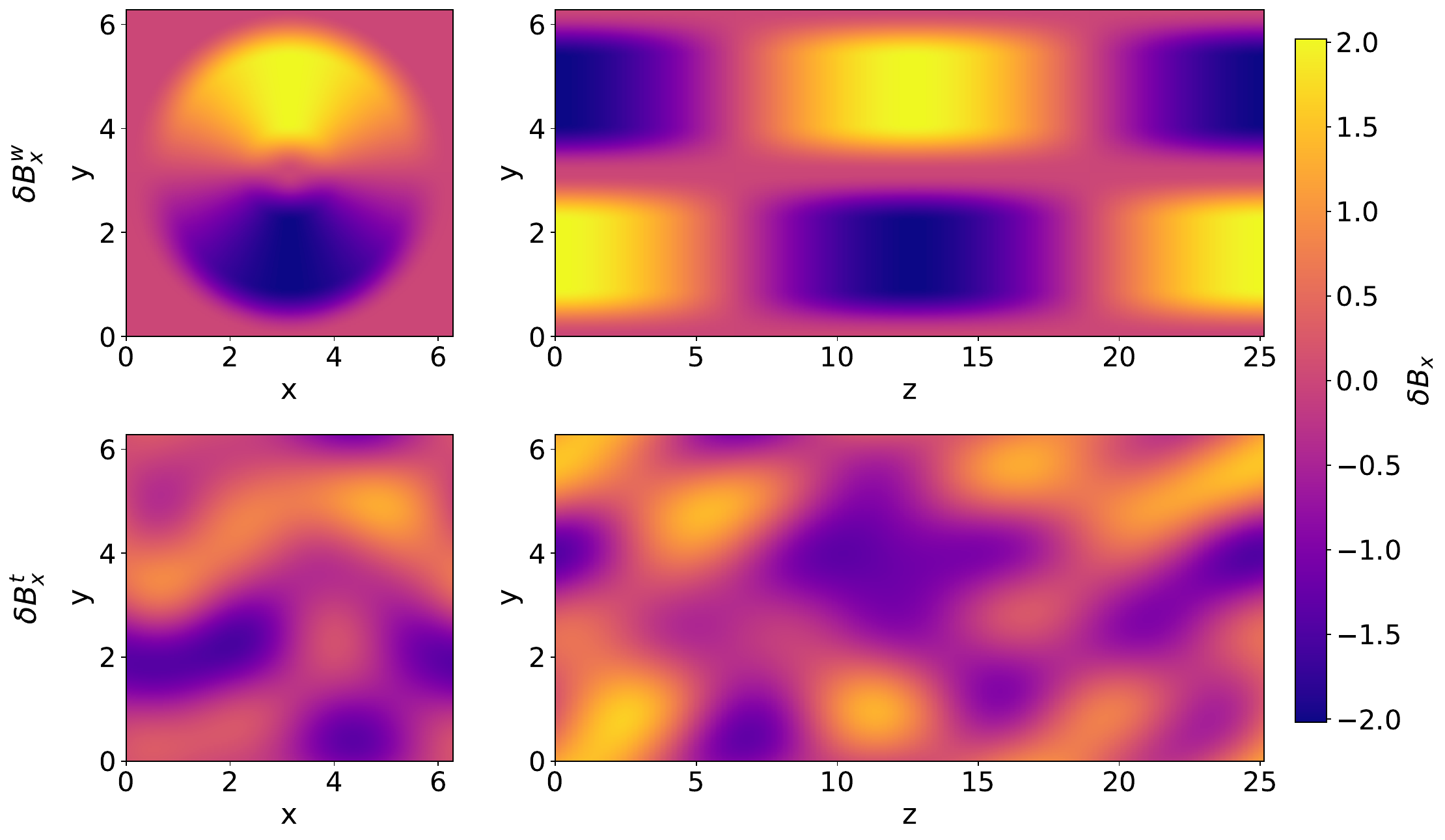}  
\caption{2D cuts of $\delta B_x$ in the $xy$ plane (left column) and $yz$ plane (right columns) at $t=0$ for the wave component (top row) and the turbulent one (bottom row).}
\label{f:IC}
\end{figure}

\noindent Spatial variations of density $\rho_0$ -- and, to a lower extent, of $B_0$ -- give rise to a nonuniform profile of the local Alfvén velocity $c_A(r)=B_0(r)/\sqrt{\rho_0(r)}$ (in code units). In particular, $c_A$ in the loop interior is lower by a factor $\simeq 0.7$ than in the loop exterior.
Those features are illustrated in Fig. \ref{f:profile0.5}, where 1D profiles of density, temperature and Alfvén velocity are shown. These profiles are taken along a line parallel to the $y$ direction intersecting the loop symmetry axis.  
The vertical dashed lines indicate the loop boundaries, which we conventionally fix at locations where $\rho=1.5$, i.e., $150$ \% of $\rho_{\rm ext}$.
\noindent The initial perturbation is made up of two components. The first component is a torsional Alfvén wave, in which the magnetic field perturbation has the form $\delta \mathbf{B}^{\rm w}(r,z)=f(r) \cos(k_z z) \hat{\boldsymbol{\theta}}$, where $f(r)$ gives the radial profile of the perturbation, $k_z=2\pi/L_{||}$ is the longitudinal wave vector corresponding to a wavelength equal to the loop length $L_{||}$, and $\hat{\boldsymbol{\theta}}$ is the unit vector in the azimuthal direction. In particular, it is $f(r)=1$ for $a\pi/2 \le r \le \pi - a\pi/2$, with $a=0.6$, while $f(r)$ smoothly goes to zero both for $r=0$ and $r=\pi$ (Paper 1). This gives a perturbation that vanishes both at the loop axis and at the domain's lateral edges.  The second component is a turbulent perturbation where the magnetic field $\delta \mathbf{B}^{\rm t}=\delta B_x^{\rm t} \hat{\mathbf{x}}+\delta B_y^{\rm t} \hat{\mathbf{y}}=\nabla \times \delta \mathbf{A}^{\rm t}$ is given by a random superposition of Fourier modes, each polarized in the transverse $xy$ plane and characterized by a 3D wave vector $\mathbf{k}$. The associated vector potential $\delta \mathbf{A}^{\rm t}$ has a power-law isotropic spectrum: $|\delta \hat{\mathbf{A}}(\mathbf{k})| \propto k^{-1.5}$ (where $\delta \hat{\mathbf{A}}(\mathbf{k})$ are the Fourier components of $\delta \mathbf{A}^{\rm t}$), including only large-scale Fourier modes $\sqrt{n_x^2 + n_y^2 + n_z^2} \le 3$, with $(n_x,n_y,n_z)$ the wave number for each direction (Paper 1). 2D visualizations of the $x$ components of $\delta B^w$ and $\delta B^t$ are shown in Fig.\ref{f:IC}.\\
\noindent
The torsional wave and the turbulent perturbation are superposed to form the initial magnetic perturbation, according to:
\begin{equation}\label{deltaB}
\delta \mathbf{B} = B_1 \frac{\alpha \delta \mathbf{B}^{\rm t}+(1-\alpha)\delta \mathbf{B}^{\rm w}}
{\left[ \langle |\alpha \delta \mathbf{B}^{\rm t}+(1-\alpha)\delta \mathbf{B}^{\rm w}|^2 \rangle \right]^{1/2}}
\end{equation}
The associated magnetic energy per volume unit is $E_{\rm M} = \frac{1}{V}\int_D |\delta \mathbf{B}|^2/2\, dV = B_1^2/2$, where $V$ is the volume of the domain $D$. Therefore, the parameter $B_1$ represents the amplitude of the magnetic perturbation. 
The parameter $\alpha$ is chosen within the interval $0\le \alpha \le 1$. It gives a measure of the relative importance of turbulence to waves. When $\alpha=0$ the magnetic perturbation includes only the contribution of the torsional Alfvén wave, while $\delta \mathbf{B}$ is due only to the turbulent component for $\alpha=1$. Intermediate values of $\alpha$ correspond to a mixture of the two components. We notice that perturbation amplitude and the associated magnetic energy depend only on $B_1$, regardless of the value chosen for parameter $\alpha$.\\
The initial velocity perturbation has the form:
\begin{equation}\label{deltav}
\delta \mathbf{v} = \frac{B_1}{\sqrt{\rho_0(r)}} \frac{\alpha \delta \mathbf{u}^{\rm t}+(1-\alpha)\delta \mathbf{u}^{\rm w}}
{\left[ \langle |\alpha \delta \mathbf{u}^{\rm t}+(1-\alpha)\delta \mathbf{u}^{\rm w}|^2 \rangle \right]^{1/2}}
\end{equation}
where $\delta \mathbf{u}^{\rm w}=-\delta \mathbf{B}^{\rm w}$ is the contribution due to the torsional Alfvén wave, while $\delta \mathbf{u}^{\rm t}$ is the turbulent component. The latter quantity has the same form as $\delta \mathbf{B}^{\rm t}$, except for the random phases of the Fourier components that are chosen differently for $\delta \mathbf{u}^{\rm t}$ and $\delta \mathbf{B}^{\rm t}$. Therefore, $\delta \mathbf{u}^{\rm t}$ and $\delta \mathbf{B}^{\rm t}$ have the same spectrum but are not spatially correlated. This gives a low value of the cross-helicity associated with the turbulent component $H_c^t=\int_D\delta\mathbf{v}^t\cdot\delta\mathbf{B}^tdV$, where $\delta\mathbf{v}^t$ corresponds to $\delta\mathbf{v}$ calculated for $\alpha=1$. In particular, it is $H_c^t \sim 0.01 E_{\rm M}^t$, where $E_{\rm M}^t = \frac{1}{V}\int |\delta \mathbf{B}^t|^2/2\, dV$ is the magnetic energy per volume unit associated with the turbulent perturbation. This corresponds to turbulent perturbations formed by fluctuations propagating both parallel and antiparallel to $\mathbf{B}_0$ with comparable amplitudes (Paper 1). This feature is essential to the development of the nonlinear energy cascade in the spectral space. We notice that, for $\alpha=0$, it is $\delta \mathbf{v}=-\delta \mathbf{B}/\sqrt{\rho_0}$, as for a torsional Alfv\'en wave. Moreover, for any value of $\alpha$, the kinetic energy per volume unit is $E_{\rm K}=\frac{1}{V}\int_D \rho_0 |\delta \mathbf{v}|^2/2\, dV = B_1^2/2$. Therefore, it is $E_{\rm K}=E_{\rm M}$, as for Alfvénic fluctuations, regardless of the value of the parameter $\alpha$. Finally, the initial density and temperature fluctuations are vanishing: $\delta \rho = \delta T =0$.
\begin{table*}[ht] \centering
\setlength\tabcolsep{5 pt}
\begin{tabular}{c c c c c c c c p{1cm}}
\hline
\hline
Run & $B_1$ & $\alpha$ & $N_{\perp}$ & $N_{\parallel}$ & $\nu=\eta=\kappa=\zeta$ & $t_d$ \\
\hline
\hline
		\rule{0pt}{2ex} A1 & 0.0 & 0.02 & 512 & 128 &  $10^{-5}$ & 270\\
        \rule{0pt}{2ex} A2 & 0.25 & 0.02 & 512 & 128 & $10^{-5}$ & 270\\
		\rule{0pt}{2ex} A3 & 0.5 & 0.02 & 512 & 128 & $10^{-5}$ & 210\\
        \rule{0pt}{2ex} A4 & 1.0 & 0.02 & 512 & 128 & $10^{-5}$ & 150\\
        \hline
        \rule{0pt}{2ex} B1 & 0.5 & 0.01 & 512 & 128 & $10^{-5}$ & 260\\
		\rule{0pt}{2ex} B2 & 0.5 & 0.03 & 512 & 128 & $2\cdot 10^{-5}$ & 150\\
        \hline
		\hline
\end{tabular}
\caption{Table with the performed runs and their parameters. $B_1$ is the perturbation amplitude. $N_{\perp}$ and $N_{\parallel}$ the number of grid points in the perpendicular and parallel direction with respect to the magnetic field $\mathbf{B}$ respectively. Parameters $\nu$, $\zeta$, $\eta$ and $\kappa$ are defined in Sect. \ref{Subsec.equations}. $\alpha$ regulates the relative amplitude of the turbulent component in the initial condition (Eq.s (\ref{deltaB}) and (\ref{deltav})). Finally, $t_d$ is the time at which the dissipation rate $W$ assumes its maximum value.}
\label{Tab1} 
\end{table*}

\subsection{Simulations parameters}\label{Sec:simparam}
We have chosen the following values of parameters describing the modeled physical system: 
\\\\
\noindent
(i) the domain aspect ratio is $\Lambda=L_{||}/L_{\perp}=4$. 
\\\\
\noindent
(ii) $\tilde{L}=1/(2\pi)\cdot 4\cdot10^4$ km, corresponding to a domain transverse size $\ell_\perp=2\pi\tilde{L}=10^4$ km, a diameter of the loop $d_\perp \simeq \ell_\perp/2=5\cdot10^3$ km and a loop length $\ell_{||}=L_{||}\tilde{L}=\Lambda \ell_\perp = 4\cdot10^4$ km.
The above value of the loop diameter $\ell_\perp$ is indeed quite high, and is due to the relatively low value of the aspect ratio $\Lambda$. However, here we are interested in describing a low-amplitude regime where phase mixing and nonlinear cascade work with comparable efficiencies to generate small scales (Paper 1). Such a low value of $\Lambda$ - though not fully realistic - allows us to describe that regime. We plan to study different regions of the parameter space in a future work.
\\\\
\noindent
(iii) $\tilde{T}=10^6$ K; this value approximately gives the plasma temperature, being the dimensionless temperature $T\simeq 1$. The value of $\tilde{T}$ corresponds to a warm loop, e.g., cooler than a typical active-region core loop \cite{reale_14}. However, in our model compressive fluctuations play a minor role; therefore, we expect that our results do not sensibly depend on the value of $\tilde{T}$, provided that condition $\beta \ll 1$ is fulfilled.
\\\\
\noindent
(iv) We set the plasma beta to $\beta=0.01$, a typical value for the Solar corona. This corresponds to a magnetic field of $\tilde{B}=11$ G.
\\\\
\noindent
(v) $\tilde{\rho}=\tilde{n}m_p= 1.67\cdot10^{-15}$ g cm$^{-3}$, where the typical number density is $\tilde{n}=10^{9}$ cm$^{-3}$ and a hydrogen plasma has been assumed. Therefore, the number density varies between $1\cdot10^{9}$ cm$^{-3}$ in the loop exterior and $2\cdot10^{9}$ cm$^{-3}$ in the loop interior
\\\\
\noindent
(vi) $\tilde{c}_A=\tilde{B}/\sqrt{4\pi\tilde{\rho}}= 759$ km s$^{-1}$, that gives an Alfvén velocity varying between $759$ km s$^{-1}$ at the loop exterior and $537$ km s$^{-1}$ at the loop interior.
\\\\
(vii) $\tilde{t}_A=\tilde{L}/\tilde{c}_A=8.28$ s, which represents the unit time of our simulations.
\\\\
\noindent
The values of the remaining parameters are given in Table \ref{Tab1} for the various numerical runs.  In Runs A1 to A4 the perturbation amplitude is fixed at the value $B_1=0.02$, while the parameter regulating the relative amplitude of the turbulent component is varied between $\alpha=0$ (torsional Alfvén wave only) and $\alpha=1$ (turbulent perturbation only). In Runs B1 and B2 the amplitude is varied between $B_1=0.01$ and $B_1=0.03$ for a fixed value $\alpha=0.5$.
As a final remark, we observe that the above values of parameters are not intended to realistically represent a specific situation, such as an active-region loop or a quiet loop. Instead, we are mainly interested in studying to which extent observations made by MUSE are capable to catch the wave dynamics related both to phase mixing and to nonlinear cascade. 
The stress is more on a dynamic range and sensitivity to variations than on absolute values.
Therefore, the choice of parameters has been dictated by the need of including both dynamical mechanisms with a comparable efficiency. 
A more accurate and realistic exploration of the parameter space will be the subject of a future work.

\subsection{Estimating spectral moments}\label{esp}
In order to investigate diagnostics of coronal heating from MUSE future observations, we are interested in estimating the observables that will be measured with MUSE.
Our aim is thus to compute line intensity $I_0$, Doppler velocity $I_1$ and non-thermal broadening $I_2$, i.e., the $0^{\rm th}$ order, $1^{\rm st}$ order and $2^{\rm nd}$ order moments of the line-of-sight velocity, respectively, at a resolution compatible with MUSE spectrograph using the synthetic density, temperature and velocity obtained from our direct numerical simulations (DNS).
In order to do so we employ the FoMo code \citep{VanDoorsselaere_16}, which is able to convert the output of the simulations into the spectral moments using the line emissivity derived from the CHIANTI atomic database \citep{delzanna_21}.
First, assuming that the plasma is optically thin and that electrons and protons have the same temperature, electron density $n$ and temperature $T$ from the DNS are converted into emission per unit area $\varepsilon$ and spectral line width $\lambda$.
The specific intensity $I$ is then computed by integrating over the chosen line of sight (in our case along the $x$ direction):
\begin{equation}
I(\lambda,y'_j,z'_k) = \sum_{\rm i} \epsilon(\lambda,x'_i,y'_j,z'_k)\Delta l.
\label{i0}
\end{equation}
$\Delta l$ being the grid step in the line-of-sight direction. 
Assuming that the synthetic $I(\lambda)$ follows a Gaussian distribution, we can derive the spectral moments from a Gaussian fit considering that:
\begin{equation}
    I_0 = \int I(\lambda)d\lambda,\\\\
    I_1 = \frac{\bar\lambda-\lambda_0}{\lambda_0}c,\\\\
    I_2 = \sqrt{\sigma^2 - \sigma_{therm}^2},\\
\end{equation}
where $\bar\lambda$ is the mean obtained from the Gaussian fit, $\lambda_0$ is the rest wavelength of the considered line, $\sigma$ is the standard deviation obtained from the Gaussian fit and corresponding to the sum of thermal and non-thermal broadening and $\sigma_{therm}$ is the broadening due to thermal effect, which we assume to be equal to $\sigma_{therm}=k_BT/m_{Fe}\sim12.18$ km/s, with $m_{Fe}$ being the mass of an iron atom.
Finally, MUSE resolution is simulated by degrading the obtained spectral moments by applying to them a point spread function, ultimately resulting in a resolution of $312$ km $\times$ $312$ km.As done in \cite{robinson_23}, we employ a square resolution element as it allows to fully demonstrate the capabilities of MUSE regardless of the observation angle.

\section{Numerical results}
We performed simulations in Table \ref{Tab1} in order to capture the dynamics of the above-described fluctuations in the simplified model of a coronal loop. 
Neglecting for simplicity compressive effects, the dissipated power is approximately given by $P_d\simeq \nu \langle \omega^2\rangle + \eta\langle J^2 \rangle$, where $\mathbf{J}=\nabla \times \mathbf{B}$ and $\boldsymbol{\omega}=\nabla \times \mathbf{v}$ are the normalized current density and vorticity, respectively. In all runs we assumed $\nu=\eta$; therefore, we considered the quantity $W=P_d/\nu=\langle J^2 \rangle + \langle \omega^2 \rangle$. Since kinetic and magnetic energies are equal in the initial perturbation, we expect that $\langle \omega^2 \rangle \sim \langle J^2 \rangle$, except for a small contribution due to the dissipation of the current associated with the equilibrium magnetic field $\mathbf{B}_0(r)$.
\begin{figure}[h]
\centering
\includegraphics[width=0.45\textwidth]{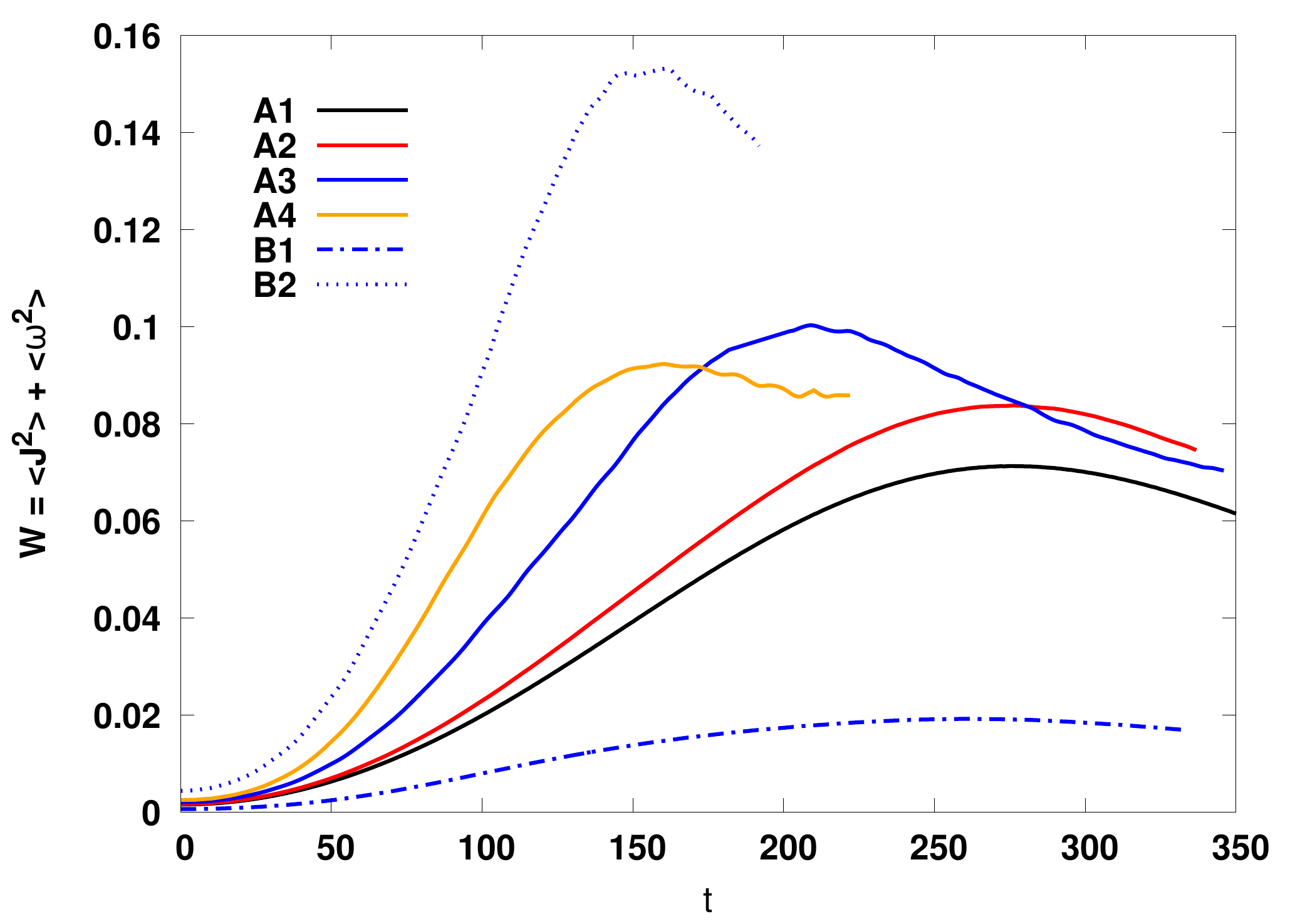}  
\caption{Time evolution of the dissipation rate W for all the runs. The time at which W reaches its maximum value, $t_d$, is indicated for each run in Table \ref{Tab1}.
}
\label{f:W}
\end{figure} 
Fig. \ref{f:W} shows the time evolution of the quantity $W$. 
In each run, during the first stage of the evolution, $W$ increases with time, indicating a progressive generation of small scales fluctuations. After reaching a maximum $W$ declines due to the dissipation of energy.
The dissipative time $t_d$ is defined as the time when $W$ reaches its maximum value. For each run, the values of $t_d$ are reported in Table \ref{Tab1}. We observe how for a fixed amplitude $B_1$, as in the case of runs A1-A4, the dissipative time $t_d$ decreases with increasing $\alpha$. 
In fact, as discussed in Paper 1, for small values of $\alpha$ the perturbation is mainly formed by the torsional Alfvén wave, whose dissipation is only due to phase mixing. In contrast, at large $\alpha$ the turbulent component is dominant; in such a case, both phase mixing and nonlinear cascade act to generate small scales, thus giving lower values of $t_d$ with respect to what obtained for small $\alpha$.\\
Morever, for larger values of $\alpha$ the background density associated with the loop structure undergoes larger distorsions (see below) with higher values of the Alfvén velocity gradient. This effect further increases the efficiency of phase mixing in generating small scales.\\
It appears also that the dissipation is higher for higher perturbation amplitude $B_1$ (run B2), therefore expecting a connection to the intensity of the global magnetic field \citep{reale_25}.
Comparing runs A3, B1 and B2 carried out using the same value of $\alpha=0.5$ and different perturbation amplitude $B_1$, we observe that the dissipative time $t_d$ decreases when increasing the amplitude $B_1$. In fact, while phase mixing is not affected by the perturbation amplitude (at least for sufficiently small amplitudes), the nonlinear cascade becomes more efficient for larger $B_1$.

\noindent In the case $\alpha=0$, corresponding to a pure torsional wave, the velocity perturbation $\propto \delta \mathbf{u}^{\rm w}/\sqrt{\rho_0} $ is cylindrically symmetric. Therefore, it preserves in time the initial cylindrical symmetry of the background density. In addition, a small increase in temperature $T$ is observed (not shown), mainly localized in the annular region of the lateral boundary of the loop, due to the dissipation of fluctuations caused by phase mixing. This temperature increase is quite small (of the order of $1 \%$), due to the small value of the perturbation amplitude, $B_1=0.02$.
\hspace*{-15cm}
\begin{figure}[h]
\includegraphics[width=0.5\textwidth]{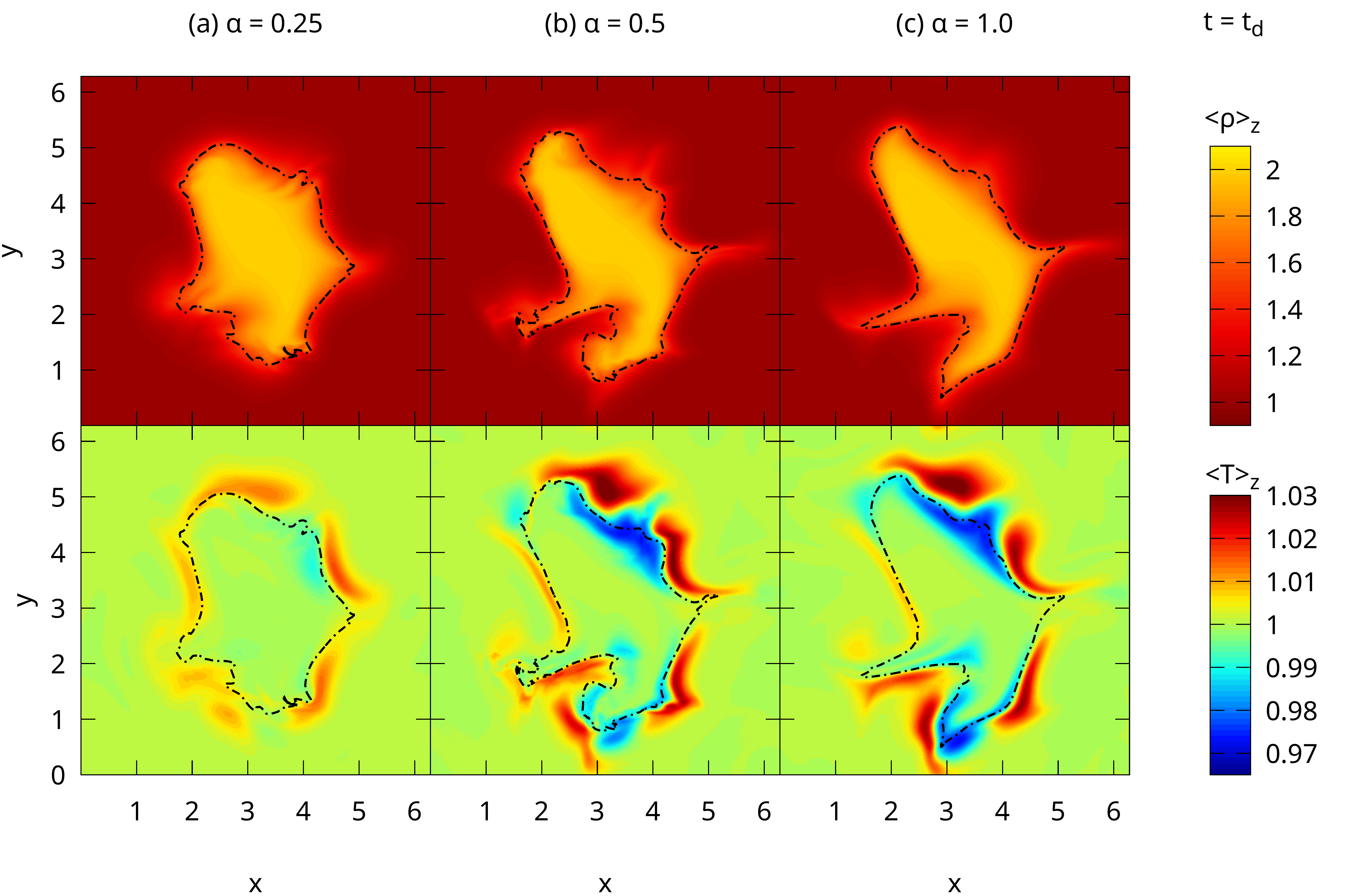}
\caption{Adimensionalized 2D density (top row) and temperature (bottom row) averaged along the direction of the loop axis, $z$, at the time dissipation reaches its peak ($t_d$) for the runs with $\alpha = 0.25$ (left), $\alpha = 0.5$ (middle), $\alpha = 1.0$ (right). The dashed black lines represent the loop boundaries.}
\label{f:rhoT}
\end{figure} 

\noindent In contrast, no particular symmetry is present in the turbulent velocity component $\delta \mathbf{u}^{\rm t}$. Therefore, when $\alpha>0$ the background density and temperature are advected by the velocity perturbation and gradually modified over time.
This is illustrated in Fig. \ref{f:rhoT} where maps of $z$-averaged $\rho$ and $T$ at $t=t_d$ in the transverse $xy$ plane are shown, for Runs A2 ($\alpha=0.25$, plot (a)), A3 ($\alpha=0.5$, plot (b)) and A4 ($\alpha=1$, plot (c)), respectively. 
In those cases, as time evolves, the shape of the loop is deformed and we observe variations in the temperature mostly alongside the boundaries of the loop.  
Comparing plots (a-c) in Fig.\ref{f:rhoT}, we also observe that deviation from a perfectly circular cross-section depends on the parameter $\alpha$: as $\alpha$ increases the deformation of the loop increases, as well. 
This indicates that the turbulent component of the perturbation is the main responsible for the loop distortion.
Moreover, temperature variations appear to be stronger with higher values of $\alpha$. This could be partially due to larger Alfvén velocity gradients, which give a faster phase mixing and therefore a more efficient dissipation.  
Finally, from Fig.\ref{f:rhoT} it is evident that maps of $z$-averaged $\rho$ in the cases with $\alpha=0.5$ and $\alpha=1.0$ are very similar.
The same holds in the maps of $z$-averaged $T$. This feature indicates that turbulence dominates the dynamics of the system even when $\alpha=0.5$.\\
\begin{figure}[h]
\includegraphics[width=0.5\textwidth]{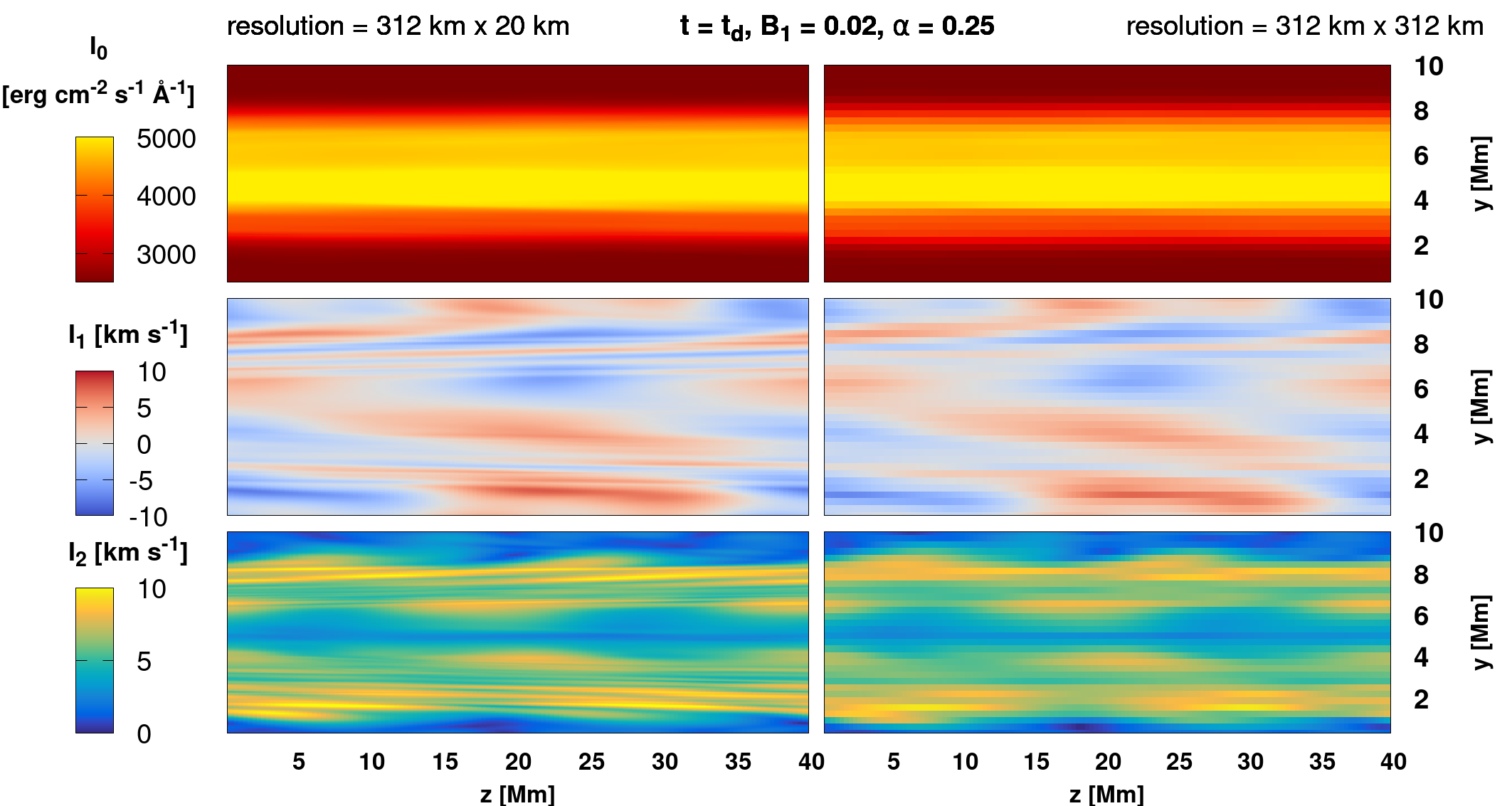}
\includegraphics[width=0.5\textwidth]{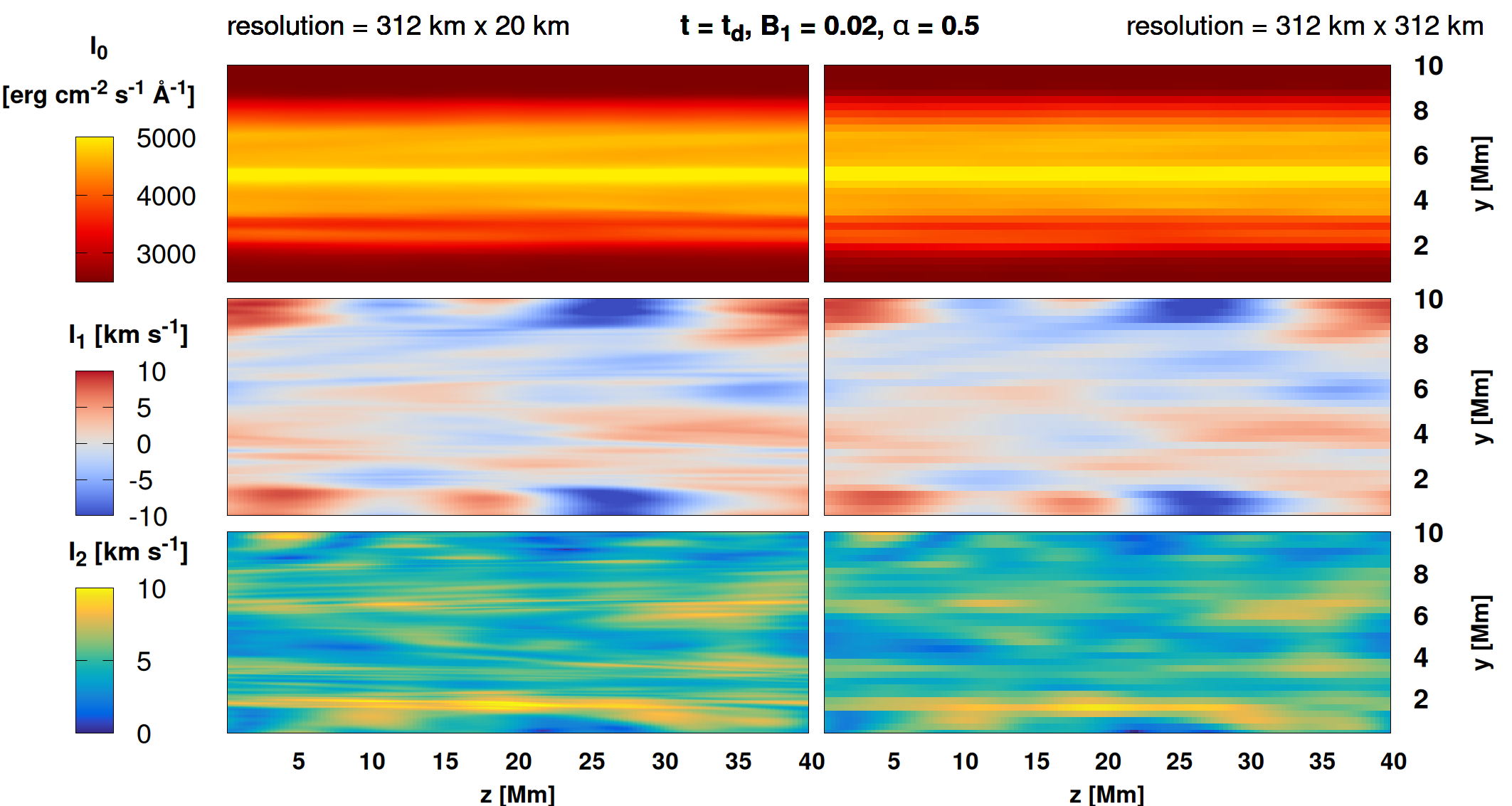}  
\includegraphics[width=0.5\textwidth]{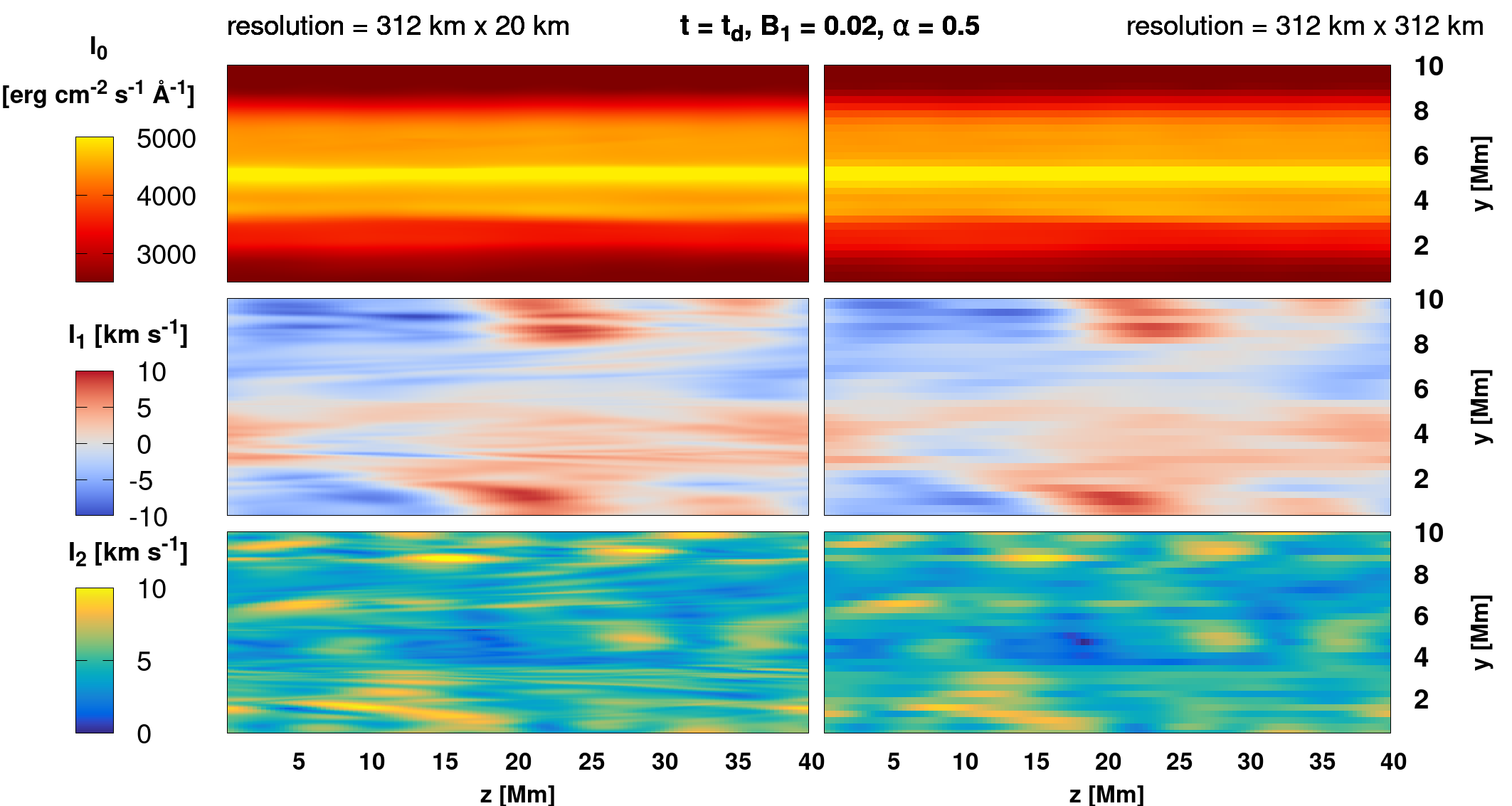}  
\caption{Spectral moments $I_0$, $I_1$ and $I_2$, computed from the synthetic fields from run A2 ($\alpha = 0.25$), A3 ($\alpha = 0.5$) and A4 ($\alpha = 1.0$). The plots in the left columns are computed using the full resolution fields obtained from the DNS, the plots in the right columns show instead the results using MUSE resolution.}
\label{f:m0.0}
\end{figure} 
\noindent We compute the spectral moments, namely, line intensity $I_0$, Doppler shift $I_1$ and non-thermal broadening $I_2$, using the FoMo software, as discussed in Sect. \ref{esp}. At this point, the electron number density $n=\rho/(\mu m_p)$, temperature $T$ and velocity $\mathbf{v}$ derived from the simulations are processed through FoMo \citep{VanDoorsselaere_16}. We assume that the line of sight is along the $x$ direction, transverse to the background magnetic field $\mathbf{B}_0$ ($z$ direction). Therefore, maps of spectral moments in the $yz$ plane have been derived, for given values of time $t$.
In this procedure we have first assumed a resolution $\Delta y \times \Delta z$ corresponding to the full resolution of DNS, i.e., $\Delta y=\ell_\perp/N_\perp\simeq 20$ km and $\Delta z=\ell_{||}/N_{||}\simeq 312$ km, where the transverse and parallel sizes of the domain are $\ell_{\perp}=10^4$ km and $\ell_{||}=4\cdot10^4$ km, respectively.
\noindent
The spatial resolution of the MUSE instrument is $0.4$ arcsec, corresponding to a resolution of $\Delta \ell_{\rm MUSE}\simeq 290$ km at a distance $d=1$ AU.
Therefore, we use FoMo to compute the spectral moments assuming a resolution of $312$ km $\times$ $312$ km, in order to have a resolution compatible with that of MUSE.
\begin{figure}[h]
\includegraphics[width=0.5\textwidth]{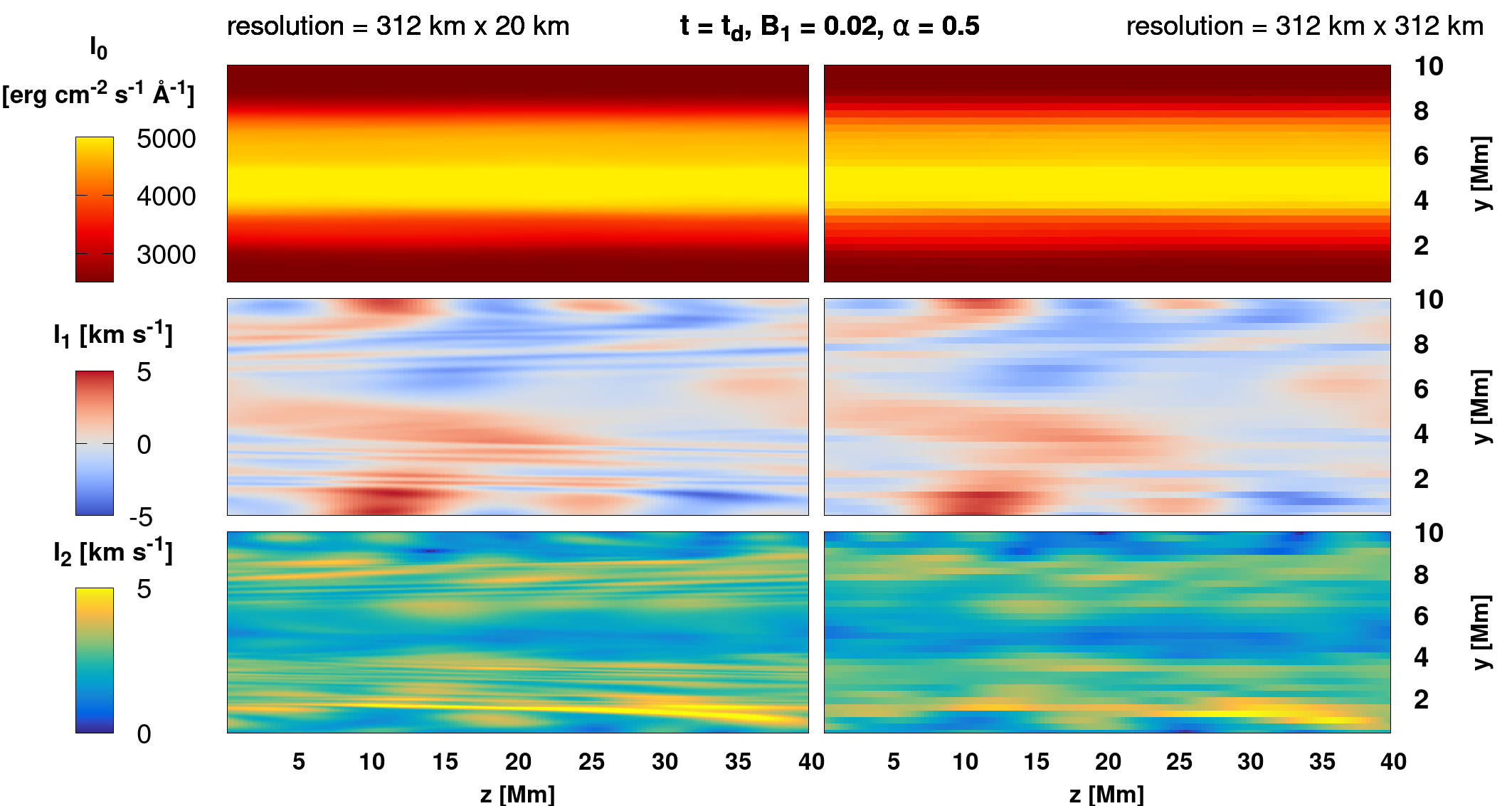}
\includegraphics[width=0.5\textwidth]{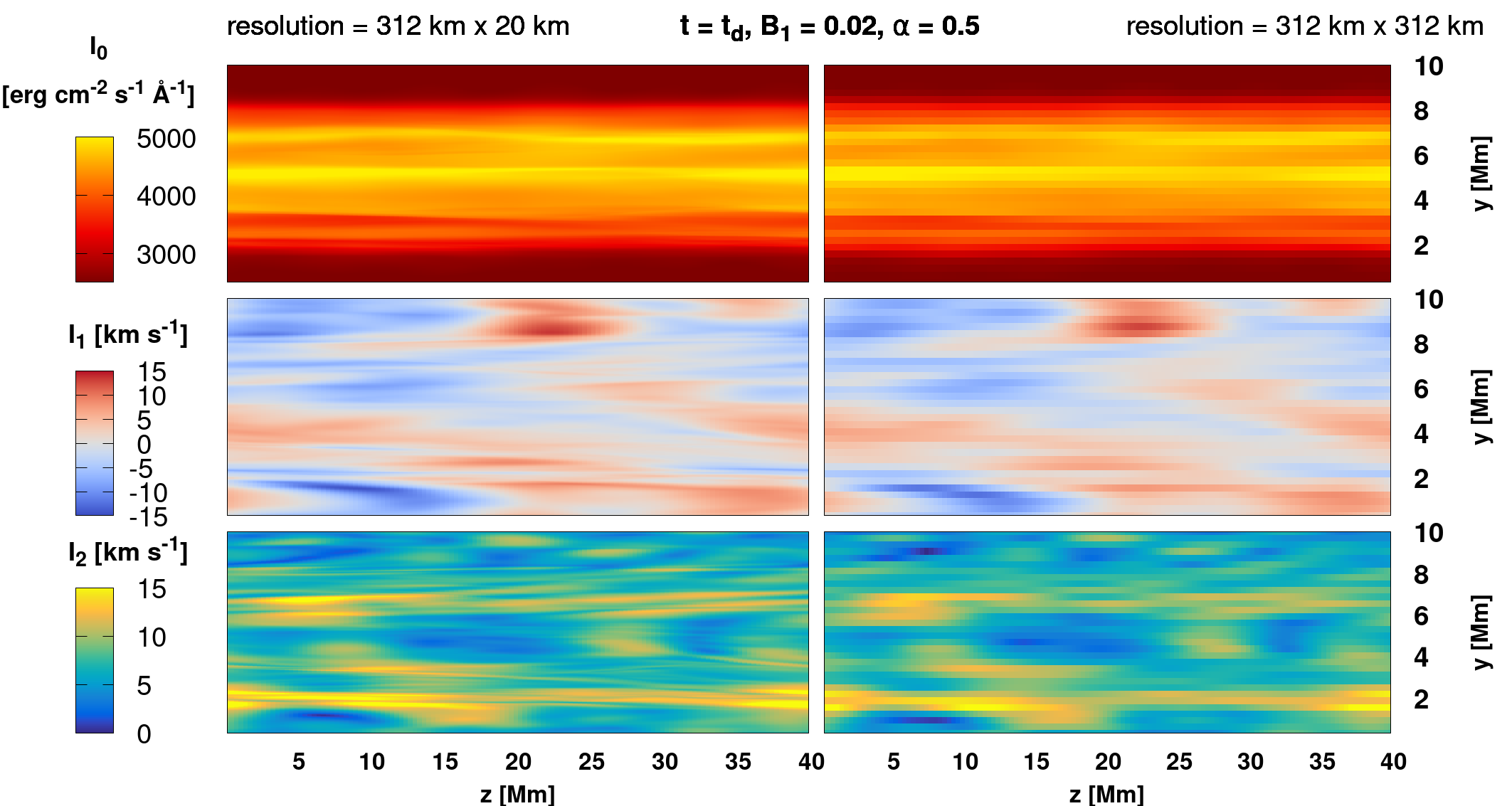}  
\caption{Spectral moments $I_0$, $I_1$ and $I_2$, computed from the synthetic fields from run B1 ($\alpha = 0.5$, $B_1=0.01$) and B2 ($\alpha = 0.5$, $B_1=0.03$). The plots in the left columns are computed using the full resolution fields obtained from the DNS, the plots in the right columns show instead the results using MUSE resolution. The color bar range has been adapted to the single run in order to highlight the features in $I_1$ and $I_2$, due to the different value of $B_1$ compared to the runs shown in Fig.\ref{f:m0.0}.}
\label{f:m0.5}
\end{figure} 
\begin{figure}[h]
\includegraphics[width=0.5\textwidth]{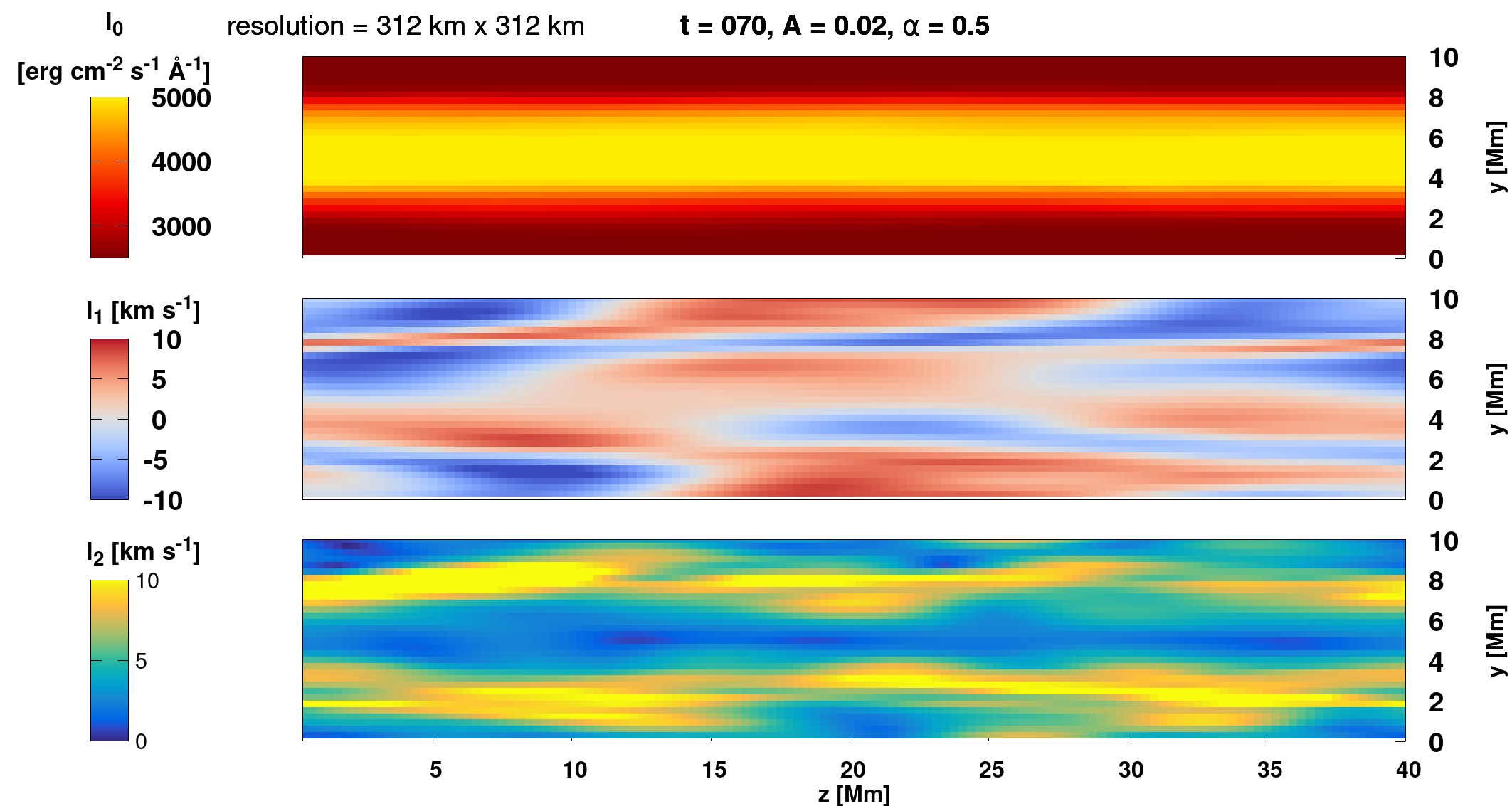}
\includegraphics[width=0.5\textwidth]{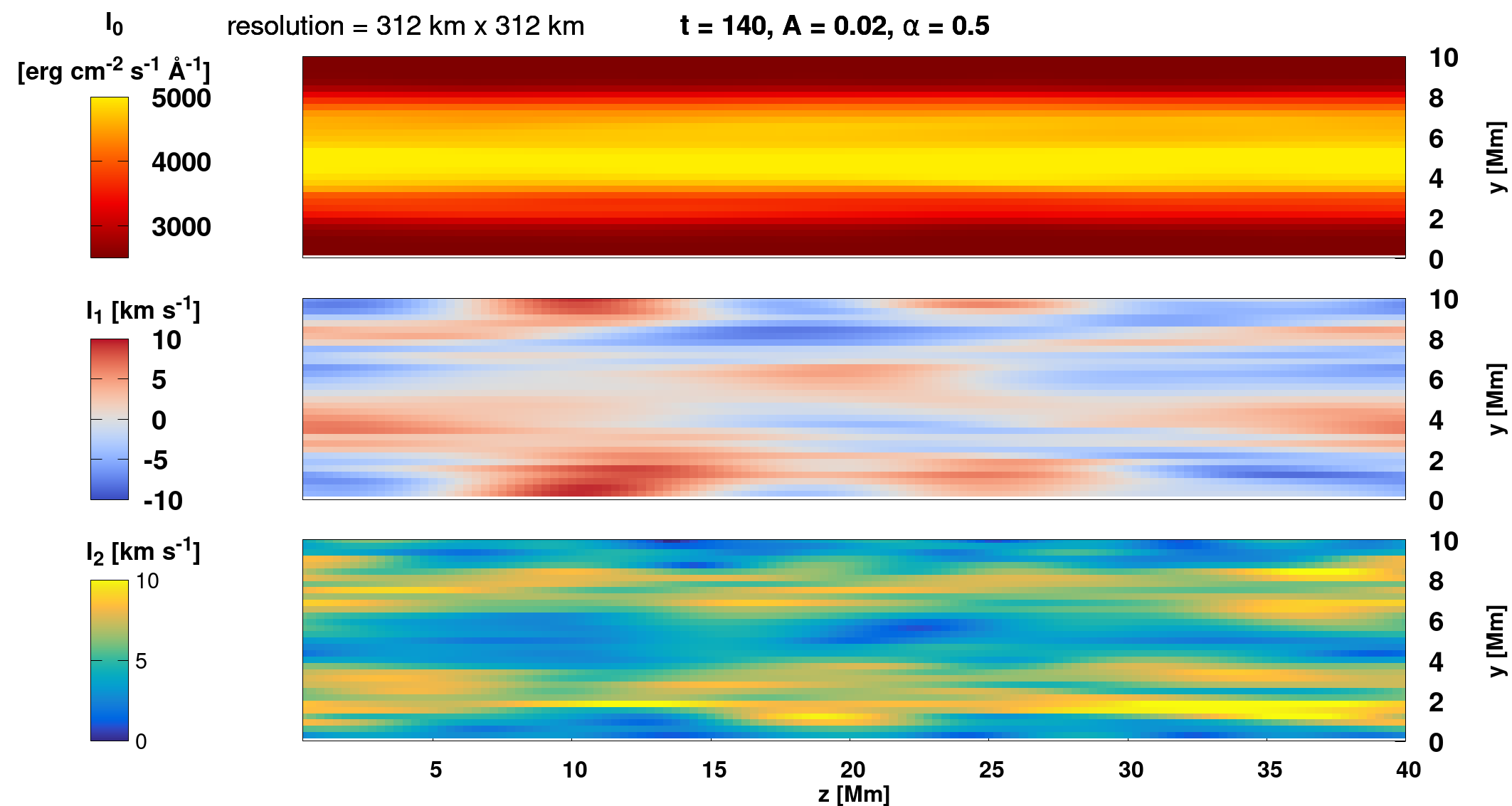}  
\includegraphics[width=0.5\textwidth]{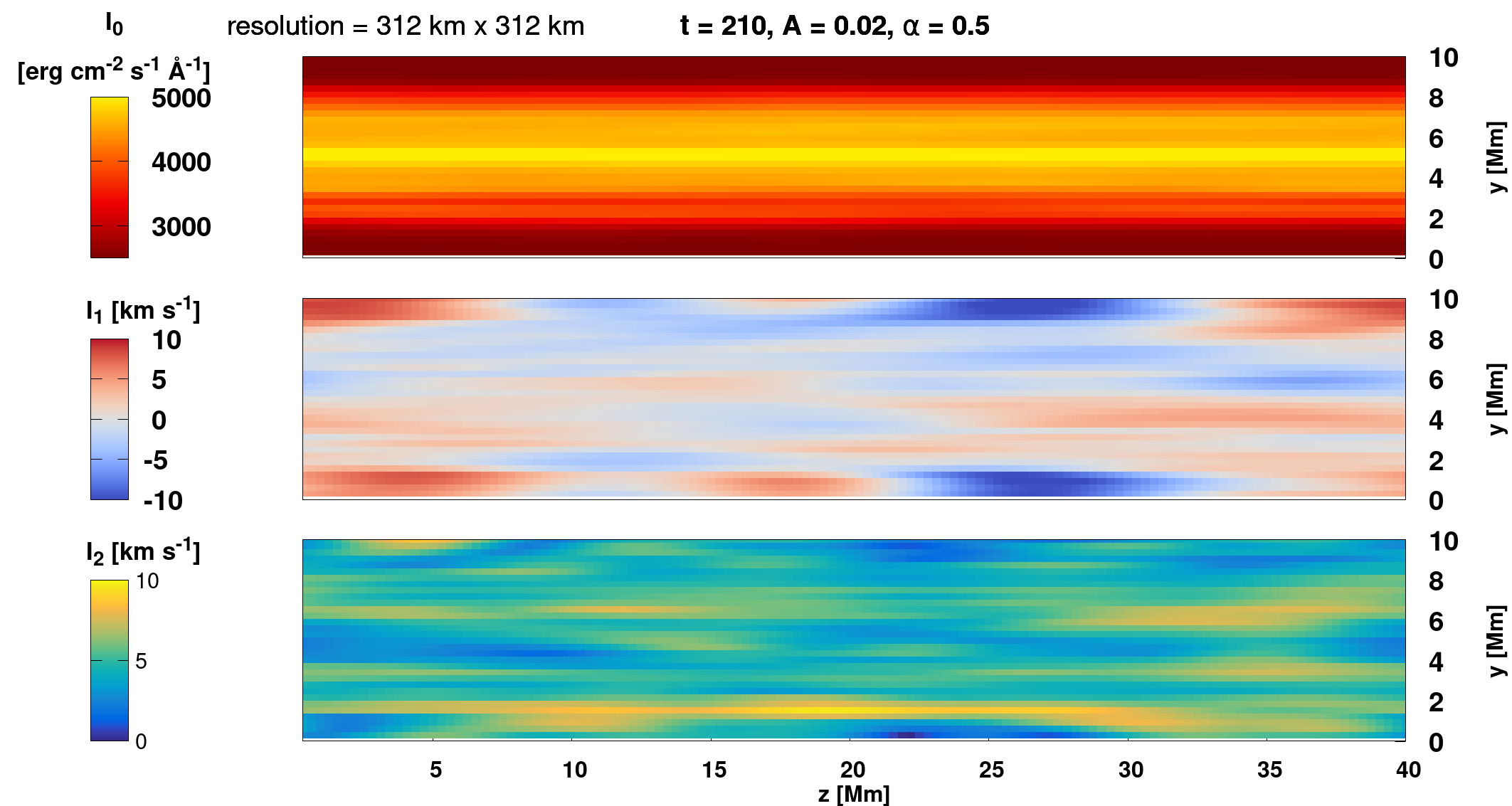}  
\caption{Time evolution of the spectral moments $I_0$, $I_1$ and $I_2$ for run A3 at MUSE resolution. The bottom panel is computed at the dissipation time $t_d$.}
\label{f:mt}
\end{figure} 
We show the comparison of the spectral moments computed with the two resolutions in Fig.\ref{f:m0.0} for runs A2, A3 and A4, in which the perturbation amplitude is fixed at the value $A=0.02$ while the parameter $\alpha$ has the values $\alpha=0.25,0.5,1$.
Spectral moments calculated at DNS resolution are plotted in the left column, while those computed at MUSE resolution are found in the right column.
All quantities in Fig. \ref{f:m0.0} are expressed in physical units.
Moreover, for each run all maps are plotted at the corresponding dissipative time $t_d$, when small scales have been fully generated.
\noindent For run A1 ($\alpha=0.0$) (not shown), where only the torsional wave is initially present, we have found that the map of $I_0$ remains unchanged throughout the run. 
This is consistent with the fact that the torsional wave alone does not modify the spatial structure of the density, keeping the shape of the loop unchanged over time. 
This is in contrast with what was found in \cite{diaz2021transition}, where they observe a transition from a wave-dominated dynamics to a turbulence-dominated dynamics after the development of Kelvin-Helmholtz instabilities, leading to the deformation of the loop shape. In our case, we believe the absence of such effect to be due to the low amplitudes $A_1$.
Maps of spectral moments relative to Run A2 are shown in the six uppermost panels of Fig. \ref{f:m0.0}. In this case ($\alpha=0.25$) the initial condition is dominated by the torsional Alfvén wave. 
Despite that, the weak turbulent component of the initial perturbation acts by modifying the initial cylindrical symmetry of the density, thus generating a deformation of the loop structure (see Fig. \ref{f:rhoT}). This results in a transverse modulation of emission intensity $I_0$ that is clearly visible at time $t=t_d$ (Fig. \ref{f:m0.0}), both at full resolution (left) and at MUSE resolution (right). 
The torsional Alfvén wave propagates in the positive $z$ direction (from left to right in the figures) with an Alfvén velocity that is larger in the outer region than inside the flux tube. Therefore, it undergoes phase mixing, which generates transverse small scales localized across the loop lateral boundary. 
This is visible on maps of both the Doppler shift $I_1$ and the non-thermal broadening $I_2$, where a pattern formed by thin stripes elongated in the parallel $z$ direction is clearly visible, localized at the boundary of the loop. 
\begin{figure*}[ht!]
\center
\hspace*{0.0cm}
\includegraphics[width=0.7\textwidth]{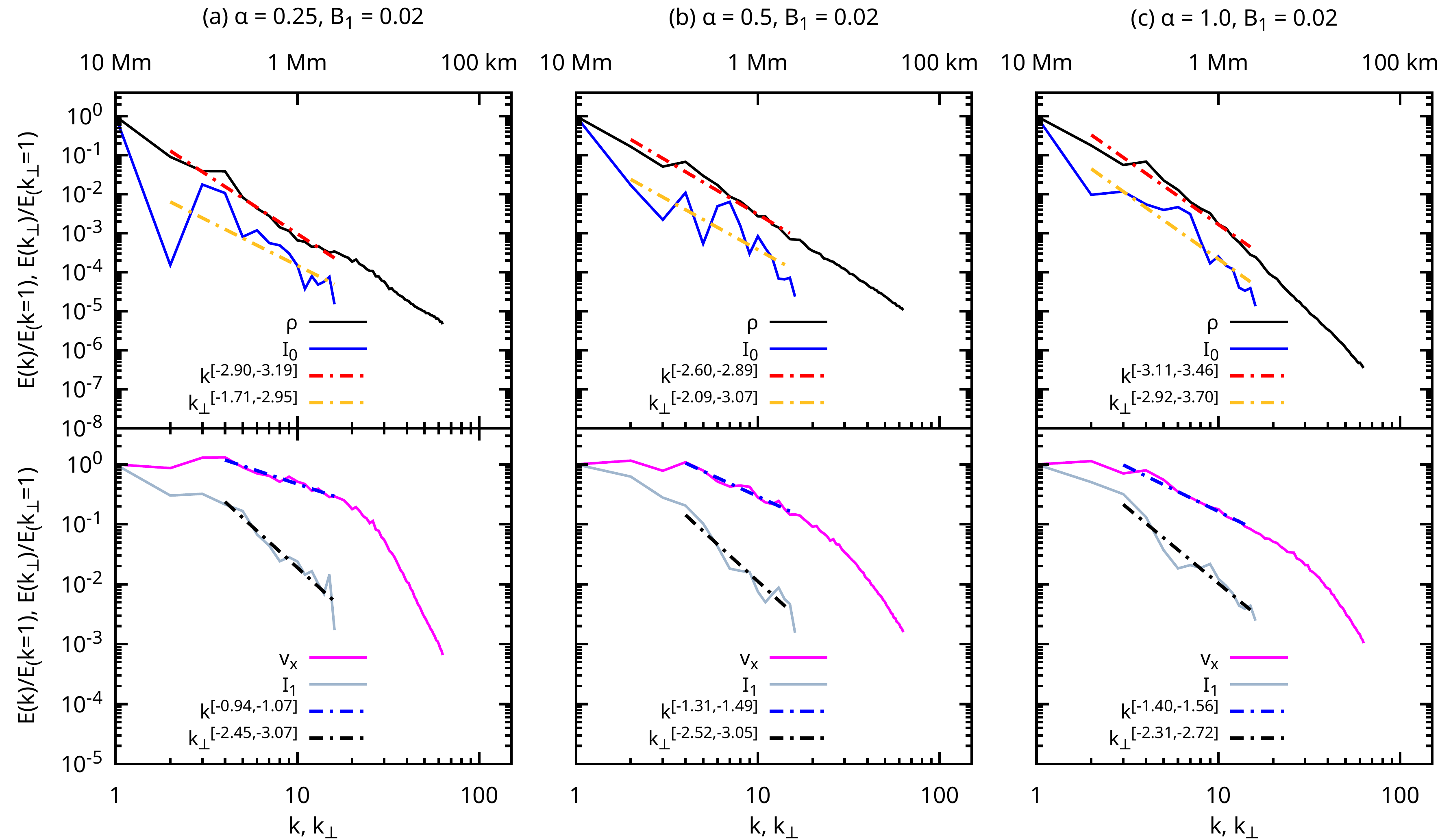}
\caption{Comparison between the power spectrum of the density 
$\rho$ and velocity along the line of sight $v_x$ computed from the $3D$ field and the power spectrum of the intensity $I_0$ and the Doppler shift $I_1$ computed from the $2D$ maps at MUSE resolution for runs A2, A3 and A4.
Each spectrum is normalized by its value at $k_{\perp}=1$ and averaged over $20 t_A$. 
Power-law fits of the power spectra are added in each plot as a reference with the range due to the uncertainty shown in the legend of each panel.}
\label{f:spectra}
\end{figure*}
\noindent Comparing the left with right panels of Fig. \ref{f:m0.0} for case $\alpha=0.25$, we see that the spatial distributions of $I_1$ and $I_2$ in the $yz$ plane obtained at the MUSE resolution (right panels) are qualitatively similar to those calculated with the full simulation resolution (left panels), although in the latter case scales smaller than at the MUSE resolution are visible. Therefore, MUSE will be able to substantially capture the pattern produced by phase mixing, at least at large scales. \\
The case $\alpha=1$ (Run A4), corresponding to an initial condition containing only the turbulent component, is illustrated in the six lowermost panels of Fig. \ref{f:m0.0}, at time $t=t_d$.
Also in this case, the density filamentation induced by the turbulent perturbation is clearly visible in the maps of $I_0$. 
The formation of small scales can be observed in the maps of $I_1$ and $I_2$. Comparing with what observed for smaller values of $\alpha$, in the case $\alpha=1$ small scales are present in a large portion of the spatial domain, probably due to the effect of nonlinear cascade that plays a more relevant role in the dynamics of perturbation. However, the thinnest structures in $I_1$ and $I_2$ maps are mostly localized in the region across the lateral boundary of the distorted flux tube, indicating that a form of phase mixing is active even in the purely turbulent perturbation (Paper 1). 
We observe that large-scale features of the spectral moments are well recovered even at MUSE resolution.
\begin{figure}[ht!]
\center
\hspace*{-2cm}
\includegraphics[width=0.7\textwidth]{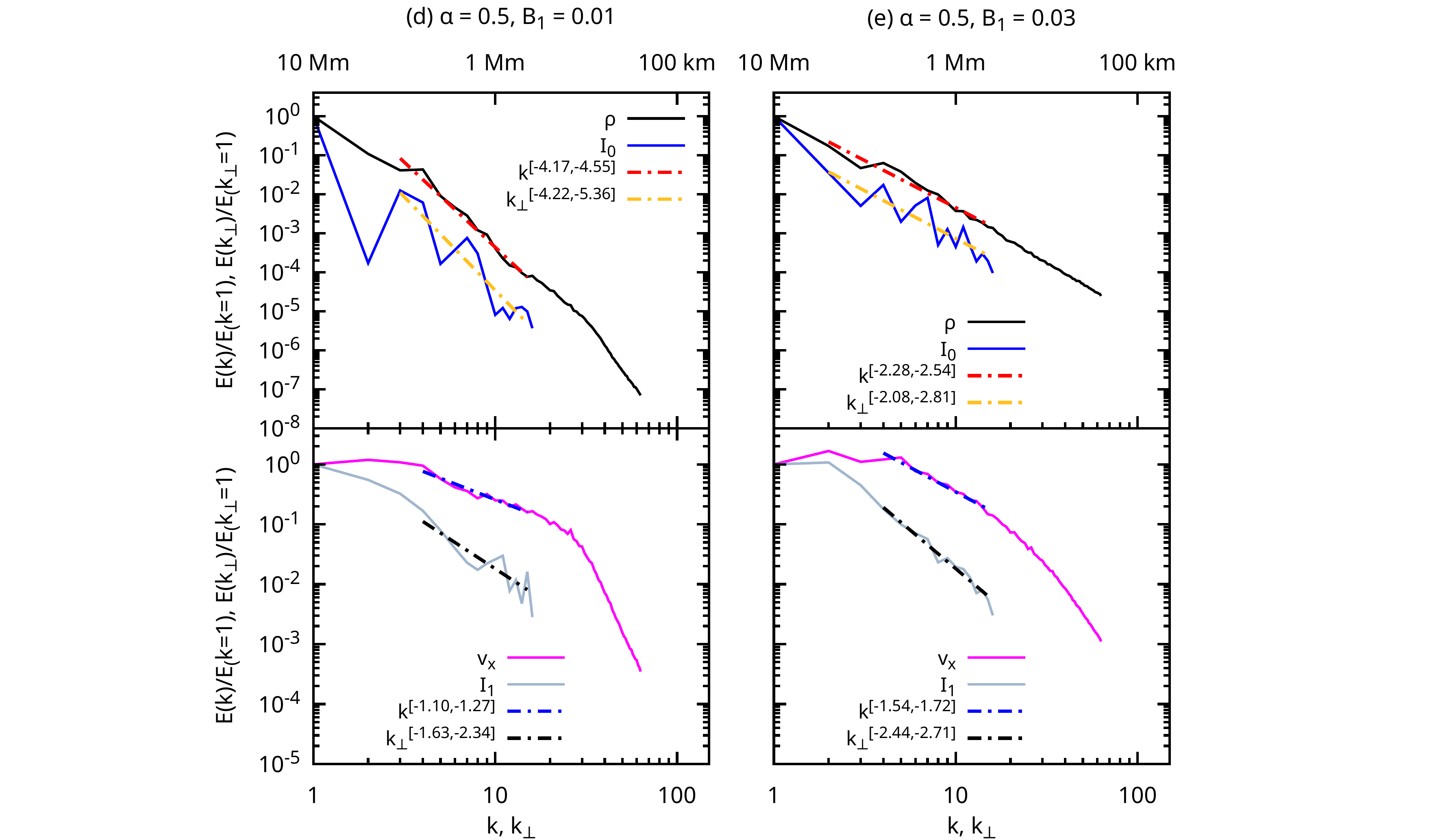}
\caption{Comparison between the power spectrum of the density $\rho$ and velocity along the line of sight $v_x$ computed from the $3D$ field and the power spectrum of the intensity $I_0$ and the Doppler shift $I_1$ computed from the $2D$ maps at MUSE resolution for runs B1 and B2.
Each spectrum is normalized by its value at $k_{\perp}=1$ and averaged over $20 t_A$. 
Power-law fits of the power spectra are added in each plot as a reference with the range due to the uncertainty shown in the legend of each panel.}
\label{f:spectra2}
\end{figure}
\noindent The time evolution of the spectral moments is shown in Fig.\ref{f:mt} for run A2.
As observed from Fig.\ref{f:rhoT}, the loop is deformed in presence of the turbulent component (i.e. $\alpha \ne 0$).
In this case, filamentation of the loop and longitudinal oscillations are observed and are more evident as $\alpha$ increases.
At early times the effects of phase mixing can be observed in both $I_1$ and $I_2$ for every value of $\alpha$. 
At later times, for high values of $\alpha$ the filamentation of the background density causes transverse small scales to be present in a larger portion of the spatial domain, when compared to the low $\alpha$ cases. 
Due to dissipation, the amplitude of $I_1$ and $I_2$ fluctuations decreases with time.
We remark that the time resolution that will be attained by MUSE is $1-4$ s \citep{depontieu_20}, that is shorter than the unit time $\tilde{t}_A=8.39$ s of our simulations. Therefore, observations carried out by MUSE should be able to follow in a detailed way a time evolution of waves and turbulence as that described by our simulations.
\noindent As done with runs A2, A3 and A4, we also compute the 2D maps of the spectral moments for runs B1 and B2, where we fixed $\alpha=0.5$ and have $B_1=0.01$ and $B_1=0.025$ respectively.
This allows us to observe the effect of the amplitude $B_1$ on the dynamics of the system.\\
\noindent The spectral moments computed at the dissipative time $t_d$ are shown in Fig.\ref{f:m0.5}.
From the comparison of $I_0$ we see how increasing $B_1$ leads to a more efficient filamentation of the loop, as seen in the bottom plot for run B2.
From the time evolution of $I_0$ (found in the supplemental material) we also observe how longitudinal oscillations of the loop are more significant for run B2.
Comparing runs B1 and B2, we see that in the latter case the dynamical ranges of variation of $I_1$ and $I_2$ are increased by a factor $\sim 3$ with respect to the former case (Fig. \ref{f:m0.5}), in accordance with the corresponding increase of the initial wave amplitude $B_1$.\\
\noindent The fine spatial resolution that will be attained by MUSE will allow for the computation of power spectra of the observables $I_0$ and $I_1$ never achieved before.\\
\noindent To show the capabilities of MUSE we now compute these power spectra using the 2D maps of $I_0$ and $I_1$ calculated at the MUSE resolution, and compare them with the isotropic power spectra computed from the full 3D physical quantities $\rho$ and $v_x$. In our case, $v_x$ corresponds to the line-of-sight velocity component, which enters in the determination of the Doppler shift $I_1$.
For this purpose, 2D maps of $I_0$ and $I_1$ in the $yz$ plane are Fourier transformed, obtaining complex Fourier coefficients in the reciprocal $k_yk_z$ space. Then, the squared modules of such coefficients are summed up over concentric shells, each shell being localized around a given value $k_\perp=\sqrt{k_y^2+k_z^2}$. This produces 1D power spectra that are functions of $k_\perp$.
These have been compared to the power spectra of $\rho$ and $v_x$, computed over isotropic shells of radius $k=\sqrt{k_x^2+k_y^2+k_z^2}$.
\noindent In Fig.\ref{f:spectra}-\ref{f:spectra2} we show  the power spectra of $I_0$ and $I_1$, compared with those of $\rho$ and $v_x$, respectively, for runs A2, A3 and A4 (panels a-c) and runs B1 and B2 (panels d-e). 
In order to make the comparison easier, each power spectra has been normalized to its value at $k_\perp=1$. Moreover, spectra have been averaged over a time interval $t_d - \Delta t_s/2 \le t \le t_d + \Delta t_s/2$ of amplitude $\Delta t_s = 20 t_A$ around the dissipative time $t_d$. This average procedure produces smoother spectra than those obtained considering a single instant of time, giving information on the energy distribution over spatial scales during the entire considered time interval.
On the lower horizontal axis of Fig. \ref{f:spectra}-\ref{f:spectra2} the values of $k_\perp$ in dimensionless units are reported, while the corresponding lengths -- expressed in physical units -- are reported on the upper horizontal axis.
$I_0$ and $I_1$ have been evaluated at the MUSE resolution, which is lower than the full DNS resolution. Therefore, while power spectra of $I_0$ and $I_1$ are defined on a $\sim 1$-decade-wide interval of the $k_\perp$ axis at larger scales, power spectra of $\rho$ and $v_x$ extends down to scales below the MUSE resolution. Of course, the comparison can be made only in the large-scale domain.
In such a range of scales the power spectrum of $\rho$ can be reasonably approximated by a power law $\propto k^{-h_\rho}$ for the three runs, indicated by red dashed-dotted lines in Fig. \ref{f:spectra}-\ref{f:spectra2}, upper panels. Similarly, the power spectra of emission intensity $I_0$ have been fitted by a power law $\propto k_\perp^{-h_{I_0}}$, indicated by an orange dashed-dotted line.
The range of values with the uncertainty of the spectral indexes $h_\rho$ and $h_{I_0}$ are indicated in the legend of Fig. \ref{f:spectra}-\ref{f:spectra2}.
We observe that in all the runs the value of the spectral index $h_{I_0}$ is larger but close to $h_\rho$ within the uncertainty, except for the case $\alpha=0.25$ when the accordance between $h_{I_0}$ and $h_\rho$ is borderline.
Therefore, calculating the power spectrum from maps of emission intensity derived by MUSE observations can give a reliable information on the distribution of density fluctuations inside the loop at the different spatial scales.
\noindent In a similar way, in the large-scale domain, the power spectra of $v_x$ and $I_1$ (Fig. \ref{f:spectra}-\ref{f:spectra2}, lower panels) are approximated by power laws $\propto k^{-h_{v_x}}$ (violet dashed-dotted line) and $\propto k_\perp^{-h_{I_1}}$ (black dashed-dotted line), respectively. In this case, the values of $h_{I_1}$ are systematically larger than $h_{v_x}$ for all the runs, i.e., the power spectra of $I_1$ is steeper than the spectrum of $v_x$. 
Therefore, taking into account the above difference, power spectra of Doppler shift maps could be still used to infer information about the distribution of velocity fluctuation at different spatial scales. We also notice that the power spectra of $I_1$ and $v_x$ reasonably follow power laws in the same range of scales (though with different spectral indexes).
The steepening observed for $I_1$ may be due to the fact that these spectra are build on a line of sight integrated field.
This leads to a conspicuous smoothing of fluctuations, which naturally leads to a steepening of the spectrum. \cite{pecora_24}.

\section{Summary and conclusions}
In the present paper we studied the dynamics of Alfvénic fluctuations in a simplified model of a coronal loop, using direct numerical simulations. The viscous-resistive compressible 3D MHD equations have been numerically solved in a three-periodic spatial domain, using the COHMPA numerical code \citep{Pezzi2024}. With respect to a previous study where we used hyperdiffusivity terms to control numerical stability (Paper 1), here we have considered a "standard" form for viscosity stress tensor, resistivity and thermal diffusivity. Moreover, we have included heating terms in the energy equation in order to self-consistently describe plasma heating due to viscous-resistive dissipation.
The loop has been modeled as a pressure-balanced cylindrical magnetic flux tube permeated by an axial-directed magnetic field $\mathbf{B}_0$, with the density inside the loop larger than outside. The plasma $\beta$ is of the order of $10^{-2}$, thus magnetic pressure prevails over plasma pressure. The Alfvén velocity is modulated in the radial direction due to the inhomogeneity of density and, to a lower extent, of magnetic field. This generates phase mixing in Alfvénic perturbations, which produces small scales across the background inhomogeneity.

\noindent From a theoretical point of view, propagation and dissipation of waves in coronal loops have been studied in depth in the literature \citep[e.g.,][]{Terradas08,Antolin14,Antolin19,Magyar15,Pagano17,terradas2018,karampelas_19,howson2020phase,shi_21,diaz2021transition,diaz2022transition}. Here we focused on the combined effect of phase-mixing and turbulent cascade in the generation of small scales in velocity and magnetic field perturbations, and the consequent dissipation. For this purpose, we have considered an initial perturbation formed by the superposition of two components. The first component is a torsional $m=0$ Alfvén wave, whose evolution is determined by phase mixing. Torsional motions have been observed both in the photosphere \citep{Brant88,Bonet08,Bonet10}, in the chromosphere \citep{Wedemeyer09,DePontieu_14,Tziotziou18,Tziotziou20,Dakanalis22} and in the transition region \citep{Wedemeyer12,DePontieu_14}. Those motions could induce torsional waves in overlying loops \citep{Jess09}. The second component is a turbulent Alfvénic perturbation localized only at large spatial scales, with an almost null cross-helicity, i.e., containing fluctuations that propagate both parallel and anti-parallel to $\mathbf{B}_0$. The latter feature is responsible for the activation of a nonlinear energy cascade which moves the wave energy to small scales. MHD turbulence with a homogeneous background has been considered in coronal heating models \citep{Nigro04,Malara10,vanBallegooijen17,rappazzo2017,vanBallegooijen18}. Moreover, there are indications of the presence of turbulent fluctuation in the corona \citep{Banerjee98,Singh06,Hahn13,Hahn14,Morton16,Morton19}. In our configuration, due to the presence of both the background inhomogeneity and a null cross-helicity, the two mechanisms -- phase mixing and nonlinear cascade -- are simultaneously at work. 
The relative amount of the two components is determined by the free parameter $\alpha$, which contributes to regulate the efficiency of the nonlinear cascade with respect to phase-mixing.

\noindent The dynamical time associated with phase mixing is independent of the wave amplitude but increases with decreasing spatial scale of the perturbation, and it can become very long for realistic values of the dissipative scale. Conversely, the time scale associated with the nonlinear cascade, which is of the order of the perturbation nonlinear time, can be quite long for small-amplitude perturbation, though it decreases with decreasing the spatial scale of perturbation. Thus, in a small-amplitude regime where phase mixing is initially faster than the cascade, phase mixing reduces the spatial scale of the perturbation until the nonlinear cascade activates. From this time on, the fluctuating energy is rapidly transferred to dissipative scales with a rate that is independent of dissipative coefficients. In this regime, phase mixing and the nonlinear cascade work in a synergistic way, speeding up the dissipation of the perturbation energy (Paper 1). Indeed, in our simulations we observed that the turbulent perturbation is also subject to phase mixing and phase mixing persists even when the initial cylindrical symmetry of the equilibrium structure is violated, indicating that it works also in configuration more complex than was originally envisaged \citep{Heyvaerts83}. Moreover, we obtained a dissipative time that is shorter for higher values of the parameter $\alpha$, i.e., in configurations where nonlinear cascade and phase mixing co-work to efficiently generate small scales.
This scenario is relevant for the corona, where waves have a small amplitude and dissipative coefficients are very low. For the above reasons, in our model we have chosen the parameters of the equilibrium structure (loop aspect ratio, density contrast) and the amplitude of the perturbation in a way to put the system in the above described regime (for $\alpha \sim 1$) (Paper 1). In this respect, we used a low value for the loop aspect ratio $\ell_{||}/d_\perp =8$. Though this is not fully realistic for a coronal loop, it allowed us to have a phase-mixing time shorter than the nonlinear time, which is a key feature of the regime we want to describe. We plan to extend the exploration of the parameter space to include more realistic representations of coronal loops in a future work. In this case, we expect to develop finer structures in the loop, due to the higher aspect ratio, which characteristic length may be closer to MUSE resolution.
Measurements that will be performed by the MUSE mission will give EUV spectra and context images with the highest resolution in space and time ever achieved for the transition region and corona \citep{depontieu_20,depontieu_22}. In this context, the main purpose of the present study has been to derive synthetic maps of observable quantities -- emission intensity $I_0$, Doppler shift $I_1$ and non-thermal broadening $I_2$ -- in order to verify the ability of the instrumentation aboard MUSE to observe the phenomena under consideration. In particular, the formation of small scales in Alfvénic fluctuations propagating in a loop, in a regime where both phase mixing and turbulent cascade can play a role in the fluctuation dynamics. For this purpose we used density, velocity and temperature, numerically calculated by simulations, as an input for the FoMo code \citep{VanDoorsselaere_16}. Since a temperature $T \sim 10^6$ K has been assumed in the model, the emission from the Fe IX line at 171 \AA\ has been considered. For all the runs, 2D maps of $I_0$, $I_1$ and $I_2$ have been derived at all simulation times, both at the DNS resolution and at the MUSE resolution. It comes out that the dynamical evolution of the system is clearly visible even at the MUSE spatial resolution (see the movies in the Supplemental Material). Moreover, the MUSE time resolution \citep[$1-4$ s, ][]{depontieu_20} is shorter than the unit time $\tilde{t}_A\simeq 8$ s of our simulations, indicating that the dynamics we described can in principle be followed in detail by MUSE observations. In particular, the initially smooth distribution of the emission intensity $I_0$ is gradually modified, leading to the formation of filaments parallel to $\mathbf{B}_0$. This corresponds to the time evolution of the background density that is gradually distorted, with the generation of gradients in the direction transverse to $\mathbf{B}_0$. This phenomenon is more evident for larger values of the parameter $\alpha$, i.e., when the turbulent component dominates the torsional wave. Moreover, a formation of small scales transverse to $\mathbf{B}_0$ is evident both in Doppler shift and in non-thermal broadening maps. This corresponds to small-scale generation in the velocity field due to phase mixing and nonlinear cascade. The effect of phase mixing, localized across the loop lateral boundaries, is more clearly visible for small and intermediate values of $\alpha$. However, some striation can be observed in the Doppler shift maps even for $\alpha=1$, indicating that phase mixing plays a role also in the dynamics of turbulent fluctuations, as expected. Such structures can be recognized also at the MUSE resolution.

\noindent The detailed maps of observables and the corresponding picture of small-scale formation and MUSE multi-slit design, which will allow the simultaneous data collection of coronal loops enabling the computation of power spectra from the moments, have suggested us to perform a comparison between power spectra of density and line-of-sight velocity with the corresponding power spectra of $I_0$ and $I_1$.
This, in order to check whether an analysis of power spectra of observable maps obtained from the MUSE instrument can give information about the spatial spectra of density and velocity. 
This comparison has been carried out considering spectra averaged in a time interval around at the dissipative time $t_d$, when we expect spectra to have the widest spectral extension in the wave number space. The power spectra of $I_0$ decreases with increasing $k_\perp$ approximately following a power law. The corresponding value of the spectral index is close to the spectral index found for the density power spectrum. This is verified for all the considered values of the parameter $\alpha$ and amplitudes $B_1$. Therefore, the power spectra derived from maps of emission intensity can give a reliable information about the distribution of density fluctuations at different spatial scales. 
Power spectra of $I_1$ also present power law ranges, but the corresponding spectral index is always larger (a steeper spectrum) than that of line-of-sight velocity.

\noindent As mentioned above, our simplified model presents some limitations. For instance, radiative loss terms have not been included in Eq. (\ref{MHD4}). Though in our case radiative loss would be negligible due to the low amplitude of fluctuations, it can become more relevant at larger amplitudes. For this reason we plan to add radiative loss terms in an upgraded version of the COHMPA code. In order to describe a regime where both phase mixing and nonlinear cascade play a role in the formation of small scales, we used a small value for the loop aspect ratio, which is not fully realistic. We plan to explore the parameter space in a future work, to include also more realistic representations of a coronal loop.

\newpage
\section{Acknowledgements}
The present work has been supported by the Italian Space Agency (ASI) within the Project “Partecipazione italiana alla missione NASA/MUSE” (CUP: F83C22001920005), related to the NASA mission “Multi-slit Solar Explorer (MUSE)”. 
Simulations have been performed at the Newton cluster at University of Calabria, supported by “Progetto STAR 2-PIR01 00008” (Italian Ministry of University and Research), and on the Luxembourg national supercomputer MeluXina.
The authors gratefully acknowledge the LuxProvide teams for their expert support.
FF., S.S. and F.V. acknowledge the Space It Up project funded by the Italian Space Agency, ASI, and the Ministry of University and Research, MUR, under Contract No. 2024-5-E.0-CUP No. I53D24000060005.
C.M. acknowledges the support from the ERC Advanced Grant “JETSET: Launching, propagation and emission of relativistic jets from binary mergers and across mass scales” (grant No. 884631). 
O.P. acknowledges the support of the FIS2 Starting Grant FIS-2023-00246 “PhAse-sPAce cOmplexity in turbulent nearly-reversible plasmas (PAPAO)” (CUP B53C24009610001) funded by the Italian Ministry of University and Research.
T.V.D. was supported by the C1 grant TRACEspace of Internal Funds KU Leuven, and a Senior Research Project (G088021N) of the FWO Vlaanderen. 
Furthermore, T.V.D. received financial support from the Flemish Government under the long-term structural Methusalem funding program, project SOUL: Stellar evolution in full glory, grant METH/24/012 at KU Leuven. 
The research that led to these results was subsidised by the Belgian Federal Science Policy Office through the contract B2/223/P1/CLOSE-UP. 
It is also part of the DynaSun project and has thus received funding under the Horizon Europe programme of the European Union under grant agreement (no. 101131534). Views and opinions expressed are however those of the author(s) only and do not necessarily reflect those of the European Union and therefore the European Union cannot be held responsible for them. T.V.D. would like to thank Francesco Malara, Francesco Pucci and Giuseppe Nistic\`o for their hospitality during his sabbatical stay at University of Calabria in spring 2025.
B.D.P. was supported by NASA.
\bibliography{biblio}

\end{document}